\documentclass[%
 reprint,
 amsmath,amssymb,
 aps,
]{revtex4-1}

\usepackage{graphicx}% Include figure files
\usepackage{dcolumn}% Align table columns on decimal point
\usepackage{bm}% bold math

\usepackage{soul}
\usepackage{color}
\soulregister\cite7
\soulregister\ref7
\soulregister\pageref7

\usepackage{xcolor}

\usepackage{arcs}

\begin{document}

%\preprint{APS/123-QED}

\title{Magnetic field influenced electron-impurity scattering and magnetotransport}% Force line breaks with \\

\author{Jingjing Feng}
\affiliation{Department of Physics, The University of Texas at Austin, Austin, Texas 78712, USA}

\author{Cong Xiao}
\affiliation{Department of Physics, The University of Texas at Austin, Austin, Texas 78712, USA}

\author{Yang Gao}
\affiliation{Department of Physics, The University of Texas at Austin, Austin, Texas 78712, USA}

\author{Qian Niu}
\affiliation{Department of Physics, The University of Texas at Austin, Austin, Texas 78712, USA}

\date{\today}% It is always \today, today,

\begin{abstract}
We formulate a classical theory ($\omega_c\tau \lesssim 1$ with $\omega_c$ as the cyclotron frequency and $\tau$ as the relaxation time) to study the influence of perpendicular magnetic field on the electron-impurity scattering process in the two-dimensional electron gas. To describe the curved incoming and outgoing trajectories, we introduce a general recipe based on an abstraction of the actual impurity scattering process to define the scattering parameters such as the incoming and outgoing momentum and coordinate jump. In this picture, we can conveniently describe the skew scattering and coordinate jump, which will eventually modify the Boltzmann equation. We find an anomalous Hall resistivity different from the conventional Boltzmann-Drude result and a
negative magnetoresistivity as a parabolic function of magnetic field. The origin of these results is 
analyzed. The relevance between our theory and recent simulation and
experimental works are also discussed. Our theory dominates in dilute 
impurity system where the correlation effect is negligible.

\end{abstract}

\pacs{Valid PACS appear here}% PACS, the Physics and Astronomy

\maketitle

%\tableofcontents

\section{Introduction}

Magneto-transport of two-dimensional electrons is an interesting but yet complicated topic in condensed matter physics. Its various behaviors, such as the Shubnikov-de Haas oscillation~\cite{Physics Kinetics}, quantum Hall conductance~\cite{Physics Kinetics}, linear magnetoresistance~\cite{PhysRevB.58.2788}, etc., contain a wealth of information about the underlying systems. However, one of the simplest questions in this field, i.e., how the electron transports through disordered materials under a magnetic field in the classical regime, has not been fully understood yet. 

In the classical ($\omega_c\tau \lesssim 1$) regime, the electron transport can be generally described by the Boltzmann equation~\cite{Boltzmann}. However, it has been pointed out that the Boltzmann equation has to be revised to incorporate the non-Markovian effect (also called memory effect \cite{Phys. Rev. Lett. 75 197, J. Stat. Phys. 87 5/6, PhysRevLett.89.266804, PRB2003, PRB2008, PRB2005}) resulting from either repeatedly scattering on the same impurity, or repeatedly passing through a region without scattering (the latter one is also called Corridor effect~\cite{Phys. Rev. Lett. 75 197, J. Stat. Phys. 87 5/6, PhysRevLett.89.266804, PRB2003}). In addition to the memory effect, there is an equally important issue that needs to be addressed, i.e. how the magnetic field affects a single electron-impurity scattering event. This problem has a fundamental difficulty in defining scattering parameters as the incoming and outgoing asymptotic trajectories are bent by the magnetic field. 

In this work, we introduce a general recipe based on an abstraction of the actual impurity scattering process to define scattering parameters for the single elastic impurity scattering. It yields the conventional scattering parameters in the absence of the magnetic field. More importantly, it can introduce an appropriate set of scattering parameters in the presence of magnetic field to calculate the differential cross section. Specifically, the real scattering process can be abstracted into a sudden switch between the initial asymptotic and final asymptotic trajectory. 
In this classical picture, we can conveniently describe the skew scattering~\cite{Physica.24.1958} and coordinate jump~\cite{PhysRevB.2.4559}, which will eventually modify the Boltzmann equation.
We then apply this recipe to the two-dimensional
Lorentz model ~\cite{PhysRevA.25.533} where free electrons are subject to in-plane electric field and out-of-plane magnetic field, and scattered by randomly distributed hard-disk impurities. 

We show the following results. 1) The magnetoresistivity is a negative parabolic function of magnetic field. Our result, together with the one from the previous theory of corridor effect \cite{PRB2003} yields a more accurate magnetoresistivity, closer to the numerical result \cite{PhysRevLett.89.266804}. 2) The obtained Hall coefficient becomes magnetic field-dependent, deviating from the Drude theory. For experiments, this deviation needs to be taken into account when converting the measured Hall coefficients to real electron densities. 3) The longitudinal relaxation time obtained in our theory depends on magnetic field which deviates from the Drude theory. 

This paper is organized in the following way. In Section II, we present
the general recipe to define scattering parameters for the impurity scattering, and use it to discuss the skew scattering and coordinate jump under magnetic field. The conventional Boltzmann equation is thus modified by these two mechanisms in the linear response regime~\cite{LRT}. In Section III, we solve the modified Boltzmann equation for the two-dimensional Lorentz model and derive the
anomalous Hall resistivity and negative magnetoresistivity. In Section IV we compare our result with relevant
simulations and experiments. Finally, we introduce a phenomenological method to include skew scattering into the Drude model.

\section{Classical theory of impurity scattering and electron transport under magnetic field}

In this section, we will formulate a classical theory of impurity scattering and electron transport in two-dimensional plane influenced by the external perpendicular magnetic field. Our theory only considers a single scattering event and ignore the well-studied non-Markovian and localization effect. One possible application of our theory is the electron transport in randomly distributed two-dimensional anti-dots under magnetic field. The anti-dots are geometrical holes punched into two dimensional electron gas (2DEG) on semiconductor GaAs \cite{APL.70.2309, Mirlin2001, Weiss1995, PhysRevLett.77.147}. 

Our theory requires $a\ll l<R$ (the meaning of those parameters will be clear later), which is due to the following assumptions. First, we assume a central scattering potential with $a$ being the characteristic size of the potential range. Second, our theory is developed under dilute limit of impurity concentration $n_{i} a^{2} \ll 1$ (where $n_{i}$ is impurity density), in which the localization/anti-localization effects are negligible ~\cite{Localization}. This condition also suggests that $a\ll l$ with $l$ being the mean free path, due to the fact that $n_i a^2=a/ l$. Third, from the classical regime $\omega_c \tau \lesssim 1$ (where $\omega_c$ is cyclotron frequency, and $\tau$ is relaxation time), the cyclotron radius $R$ is larger than the mean free path, i.e. $R>l$, which is the necessary condition to avoid repeated scattering at the same impurity. Summing up all the above requirement, the pre-condition of our theory is $a\ll l<R$. 

The derivation and discussion in this section are organized as follows. First, we review a critical issue in formulating our theory: as the incoming and outgoing asymptotic trajectories are bent by the magnetic field, it is not clear how to parameterize them. To resolve this issue, we introduce a general recipe to redefine the impact parameter, the incoming and outgoing momentum, and the scattering angle. Then we use them to naturally describe and calculate the skew-scattering and coordinate jump at the presence of the magnetic field, which will eventually modify the Boltzmann transport equation. 

\begin{figure}[b]
\setlength{\abovecaptionskip}{0pt}
\setlength{\belowcaptionskip}{0pt}
\scalebox{0.4}{\includegraphics*{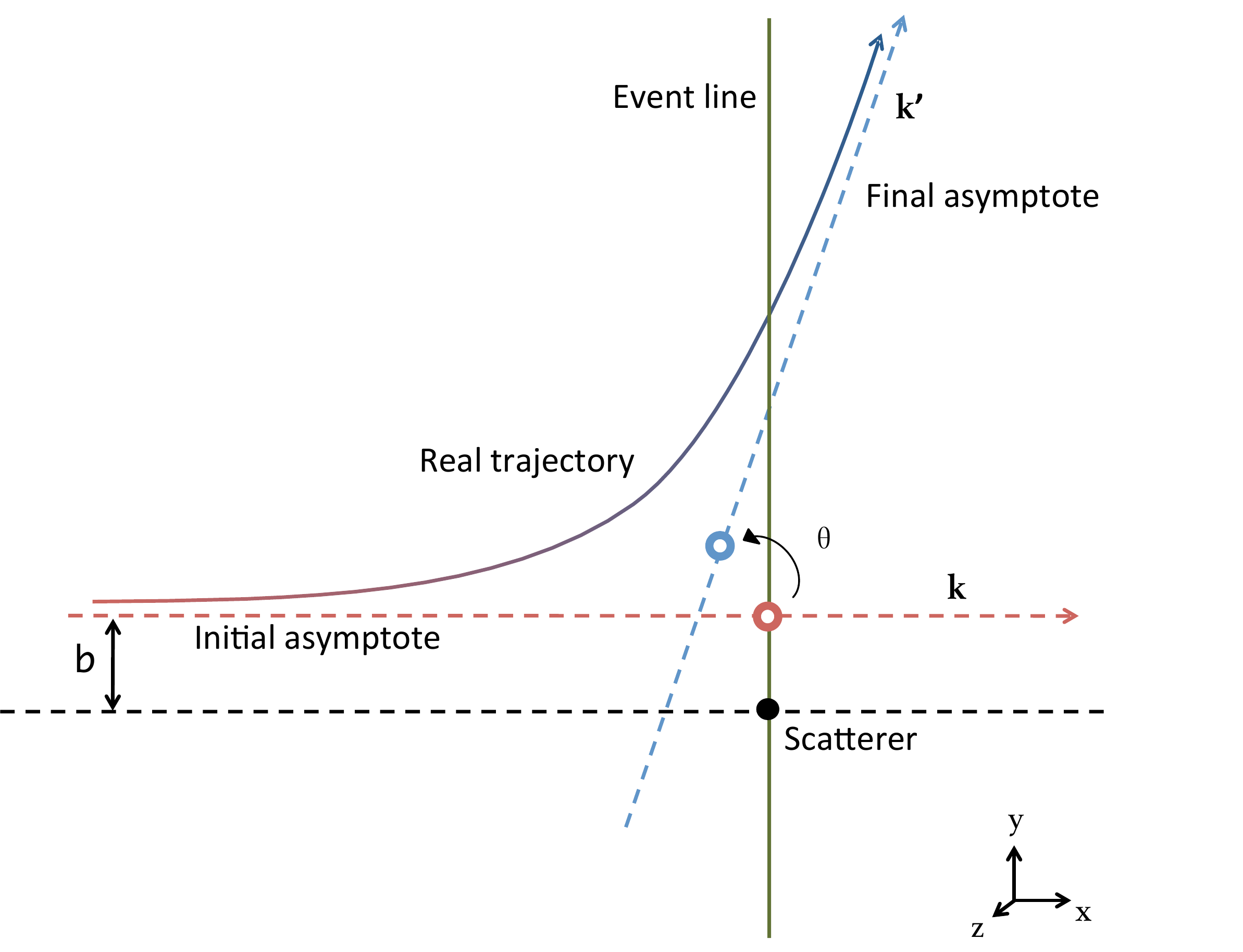}}
\caption{The illustration of a general scattering process without magnetic field. The solid curve is the real trajectory starting from color red and ending with color blue. The red and blue dashed lines are the initial and final asymptotic trajectory, respectively. The red and blue empty dot are the starting point and ending point, respectively. The green line is the event line passing through starting point and impurity center. }
\label{fig_event_line}
\end{figure}

\subsection{Abstraction of impurity scattering process}
Our final goal is to study the electron transport under magnetic field in a classical picture. Therefore, we will use the Boltzmann transport equation. It contains two parts: one describes the electron drifting between collisions driven by external forces, and the other one describes the electron scattering off impurities that leads to electronic steady states. In our situation, the drifting part is simply driven by the Lorentz force, which is well known. Therefore, we will focus on the impurity scattering under magnetic field. 

There is a critical issue in describing such a scattering process classically. To see this, we first review the conventional electron-impurity scattering in the absence of the magnetic field. In Fig.~\ref{fig_event_line}, we plot such a scattering process. The real electron trajectory is represented by the solid curve, with the arrow showing the direction of the electron motion. Then scattering parameters such as the impact parameter, the incoming and outgoing momentum, and the scattering angle are easily defined from the incoming and outgoing asymptotic trajectories, as illustrated in Fig.~\ref{fig_event_line}. In the presence of a constant out-of-plane magnetic field, however, such definition of scattering parameters does not work, because the above quantities vary in time in the asymptotic sense due to the curved incoming and outgoing asymptotic trajectories. 

To resolve this issue, we propose a recipe to define those scattering parameters generally. First, we introduce the abstraction of the impurity scattering process. It proceeds as follows: 
 we assume the scattering occurs suddenly at the time $t=0$; we then use the asymptotic trajectories as $t\rightarrow-\infty$ and $t\rightarrow \infty$  to replace the true trajectory at $t<0$ and at $t>0$, respectively. We call those imaginary trajectories as the initial and final asymptote, respectively. 
We define this method as the abstraction of the impurity scattering process, as it only keeps the essence of the scattering process, i.e. the transition from the initial asymptote to the final asymptote, and abstract the detail of the transition as a sudden switch. 

There is a degree of freedom in the above procedure. Note that even though we have restricted the scattering to occur at $t=0$, this point itself is not well defined. In other words, we have the freedom to define this artificial point. For a central scattering potential, we can fix this issue by requiring that at $t=0$ the electron reaches the point in the initial asymptote closest to the scatter. We call this point the starting point (represented by the red dot in Fig.~\ref{fig_event_line}). If the scattering potential respects the rotational symmetry, the starting point in different initial asymptotes form a straight line called the event line which marks the occurring of scattering event as illustrated in Fig.~\ref{fig_event_line}. It turns out that the event line is orthogonal to the initial asymptotes and passes the center of the scatterer. 

With the help of the abstraction of the impurity scattering process, we define the scattering parameters as follows. We define the distance between the starting point and the scattering center to be the impact parameter, the momentum at $t=0_-$ and $t=0_+$ to be the incoming and outgoing momentum, respectively, and the angle between the incoming and outgoing momentum to be the scattering angle. Those scattering parameters reduce to the conventional ones in the absence of the magnetic field, as shown in Fig.~\ref{fig_event_line}. We further define the point in the final asymptote at $t=0_+$ to be the ending point (represented by the blue dot in Fig.~\ref{fig_event_line}). This definition of scattering parameters is clearly independent of the scattering details and works for any type of the initial and final asymptotes.

Using the above concepts, the abstraction of the scattering process can be concisely stated as follows: the electron moves along the initial asymptote to the starting point, gets scattered to the ending point and finally moves away from the scatterer along the final asymptote.

\begin{figure}[b]
\setlength{\abovecaptionskip}{0pt}
\setlength{\belowcaptionskip}{0pt}
\scalebox{0.39}{\includegraphics*{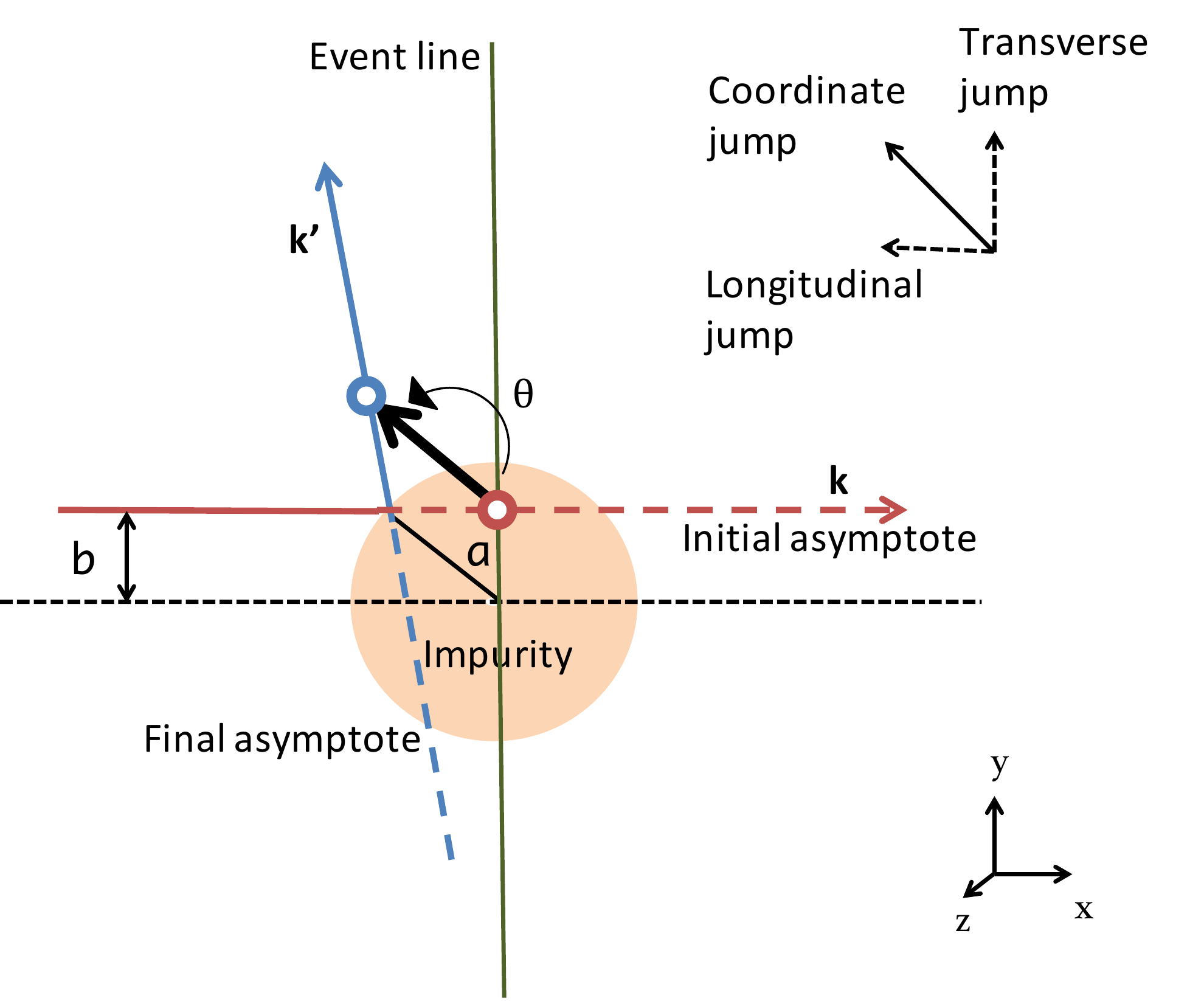}}
\caption{The illustration of the conventional hard ball scattering with no magnetic field. The red and blue empty dot are the starting point and ending point, respectively. The green line is the event line passing through starting point and impurity center. The initial asymptote and final asymptote are marked by dashed red and blue line with the incoming momentum $\mathbf{k}$, the outgoing momentum $\mathbf{k'}$, and the angle of scattering $\theta$, the impact parameter $b$. The coordinate jump can be divided into two directions, which are transverse jump and longitudinal jump. }
\label{fig_traditional hard ball}
\end{figure}

\begin{figure}[b]
\setlength{\abovecaptionskip}{0pt}
\setlength{\belowcaptionskip}{0pt}
\scalebox{0.4}{\includegraphics*{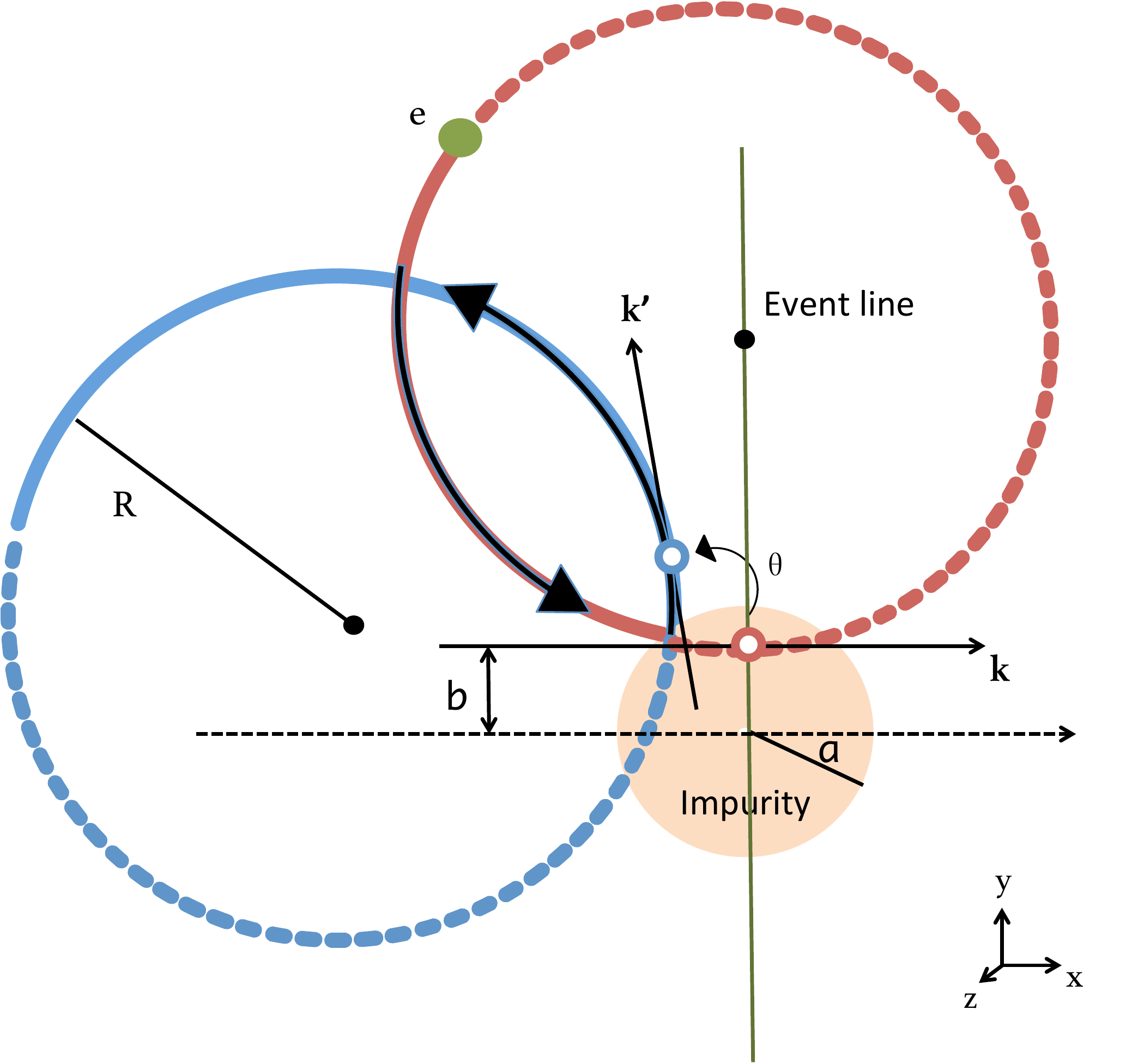}}
\caption{The illustration of electron scattering on hard disk impurity with cyclotron orbit under magnetic field. The impurity radius is $a$. The cyclotron radius is $R$. 
The red and blue solid lines are the real trajectory of the incoming and outgoing electron, respectively. The red and blue complete circle forms the initial asymptote and final asymptote. The red and blue empty dots are the starting point and ending point, respectively. The incoming momentum
$\mathbf{k}$ and the outgoing momentum $\mathbf{k}^{\prime}$ are along the tangential direction at the starting point and ending point. The angle of scattering $\theta$ is the angle between $\mathbf{k}$ and $\mathbf{k'}$. }
\label{fig_pig_picture}
\end{figure}

\subsection{Application to hard disk potential}

We first apply the abstraction of the scattering process to hard disk potential in the absence of magnetic field. By applying to this fully known case, we aims at a necessity check of the correctness of our theory. 
Consider an electron incident on a hard disk potential with straight line trajectory (Fig.~\ref{fig_traditional hard ball}). The real trajectory (solid lines) changes its direction after the electron hits the scatterer. However, the initial and final asymptote (dashed lines) can be elongated along the real trajectory and pass through the scatterer. The event line that marks the occurring of scattering event, passes through the center of scatterer and the starting point (red empty dot) on the initial asymptote. The incoming momentum $\mathbf{k}$ and outgoing momentum $\mathbf{k'}$ are defined as the starting (red empty dot) and ending point (blue empty dot) on the initial and final asymptote, respectively. 

In contrast, in the presence of magnetic field, the trajectory is bent, and we use the abstraction of the scattering process discussed in the previous subsections to define scattering parameters, as shown in Fig.~\ref{fig_pig_picture}. The incoming momentum $\mathbf{k}$ and outgoing momentum $\mathbf{k'}$ cannot be defined straightforwardly, due to the directions of the initial/final asymptote changes over time. As shown in Fig. ~\ref{fig_pig_picture}, the red and blue dashed lines are the asymptotic trajectory which completes the circular trajectory. 
The incoming $\mathbf{k}$ and outgoing $\mathbf{k'}$ are defined along the tangential direction to the initial asymptote and final asymptote at the starting point and ending point, respectively (see Fig.~\ref{fig_pig_picture}). The $\mathbf{k}$ and $\mathbf{k'}$ are rotated by the same angle in unit time. 

In the Appendix \ref{APP-E}, we demonstrate how the abstraction method can be applied to the soft potential under magnetic field. 

\begin{figure}[h]
\setlength{\abovecaptionskip}{0pt}
\setlength{\belowcaptionskip}{0pt}
\scalebox{0.38}{\includegraphics*{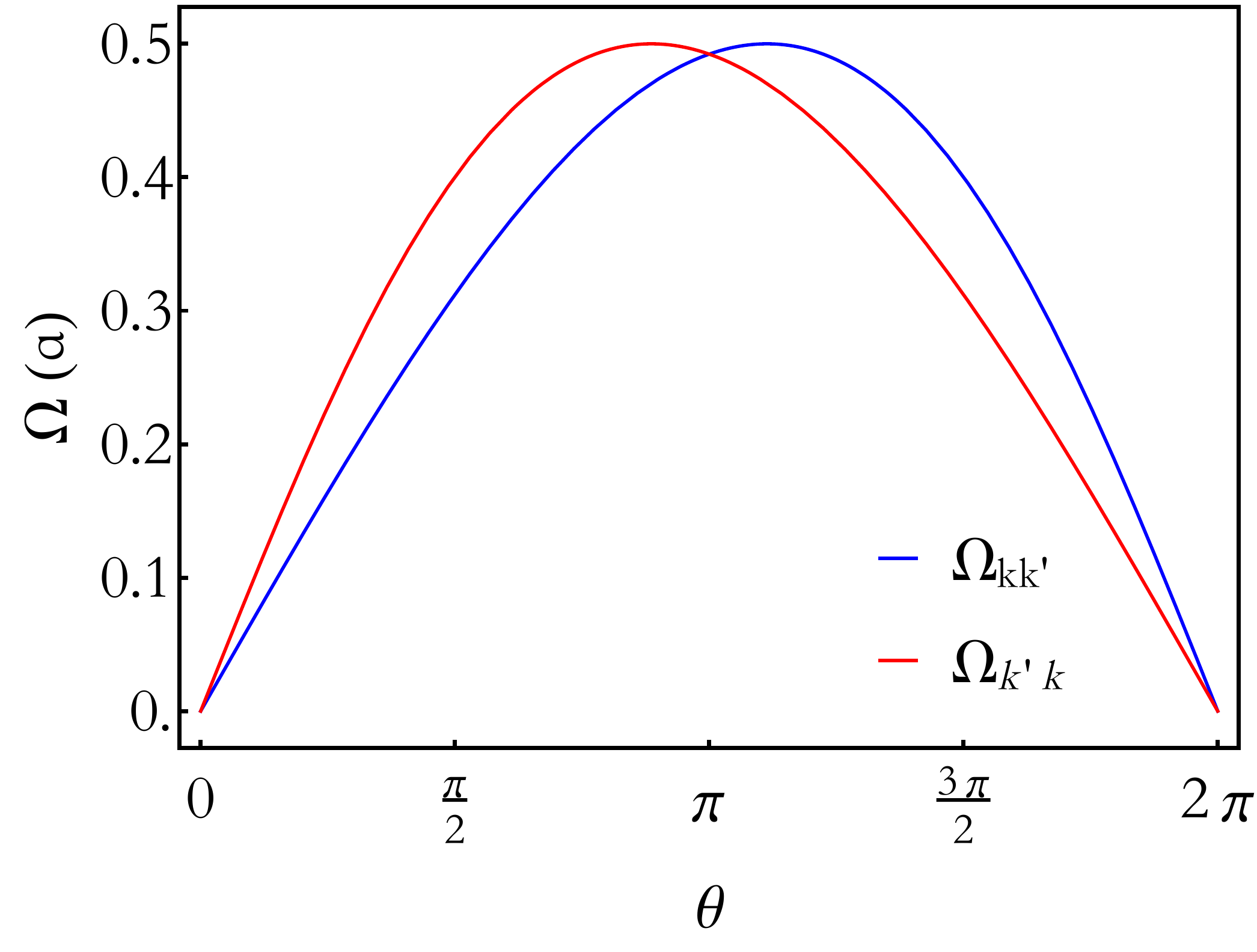}}
\caption{The plots of the differential cross section of two processes: $\mathbf{k} \rightarrow \mathbf{k'}$ and $\mathbf{k'} \rightarrow \mathbf{k}$, respectively, in the unit of impurity radius $a$. The ratio $\frac{a}{R}=0.16$. The $\Omega_{\mathbf{kk}^{\prime}}$ does not overlap with $\Omega_{\mathbf{k}^{\prime}\mathbf{k}}$, which leads to skew scattering. }
\label{fig_cross section curve}
\end{figure}

\subsection{Skew scattering under magnetic field}
In this section, we discuss the skew scattering in the classical picture. As shown in previous literatures, the antisymmetric part of the probability of scattering $W_{ \mathbf{k} \mathbf{k}^\prime}$ leads to the skew scattering~\cite{Physica.24.1958}. $W_{ \mathbf{k} \mathbf{k}^\prime}$ (probability of scattering of $\mathbf{k} \rightarrow \mathbf{k'}$ process) is related to the differential cross section as $W_{ \mathbf{k}\mathbf{k}^\prime}=n_i v_{ \mathbf{k}} \Omega_{ \mathbf{k}\mathbf{k}^\prime}$, where $n_i$ is the impurity concentration and $v_{ \mathbf{k}}$ is the electron velocity. For hard-disk potentials, the scattering is elastic, i.e. $|v_{ \mathbf{k}}|=|v_{ \mathbf{k}^\prime}|$. Therefore, a nontrivial antisymmetric part of $W_{ \mathbf{k} \mathbf{k}^\prime}$ only comes from that $\Omega_{ \mathbf{k}\mathbf{k}^\prime} \neq \Omega_{ \mathbf{k}^\prime\mathbf{k}}$. 

Using the scattering parameters shown in Fig.~\ref{fig_pig_picture}, the differential cross section is easily calculated by $\Omega_{\mathbf{k}\mathbf{k}^\prime}=\left\vert \frac{db}{d\theta}\right\vert $. 
Here we use the fact that $b$ is only a function of $\theta$ and $k=|\mathbf{k}|$ due to the rotational symmetry and the elastic nature of scattering. For two-dimensional Lorentz model, the relation between $b$ and $\theta$ and $k$ (with $R=\hbar k/(eB)$) is given by (derived in Appendix \ref{APP-A})
\begin{equation}
b(\theta,k)=-R+\sqrt{a^{2}+R^{2}+2aR\cos\frac{\theta}{2}},
\label{b}
\end{equation}
Therefore, the differential cross section reads as
\begin{equation}
\label{eq_kkprime}
\Omega_{\mathbf{kk}^{\prime}}=%
\frac{a\sin\frac{\theta}{2}}%
{2\sqrt{1+2\frac{a}{R}\cos\frac{\theta}{2}+\left(  \frac{a}{R}\right)  ^{2}%
}}.
\end{equation}

On the other hand, 
the differential cross section of the inverse process $\mathbf{k}^{\prime}\rightarrow \mathbf{k}$ is labeled by $\Omega_{\mathbf{k}^{\prime}\mathbf{k}}$, and can be calculated as follows: $\Omega_{\mathbf{k}^\prime\mathbf{k}}=\left\vert \frac{db}{d\theta}\right\vert_{\theta\rightarrow 2\pi-\theta} $. Therefore, its expression reads as
\begin{equation}
\label{eq_kprimek}
\Omega_{\mathbf{k}^{\prime}\mathbf{k}}=\frac{a\sin\frac{\theta}{2}}%
{2\sqrt{1-2\frac{a}{R}\cos\frac{\theta}{2}+\left(  \frac{a}{R}\right)  ^{2}%
}}.
\end{equation}

We plot $\Omega_{\mathbf{k}\mathbf{k}^\prime}$ and $\Omega_{\mathbf{k}^\prime \mathbf{k}}$ in Fig.~\ref{fig_cross section curve}. It shows that $\Omega_{\mathbf{k}\mathbf{k}^\prime}\neq \Omega_{\mathbf{k}^\prime \mathbf{k}}$, leading to the nontrivial skew scattering contribution to the electron transport in two-dimensional Lorentz model. In Eq. \ref{tau-perp} in Section III B and Section IV C, we will find out that only when $\Omega_{\mathbf{k}\mathbf{k}^\prime}\neq \Omega_{\mathbf{k}^\prime \mathbf{k}}$, there is $\frac{1}{\tau^{\perp}} \neq 0$ ($\frac{1}{\tau^{\perp}}$ is the reciprocal of transverse relaxation time), which is the signature of skew scattering. We further comment that the nature of the above inequivalence is a finite magnetic field, i.e. only in the limit $\mathbf{B}\rightarrow 0$, $R\rightarrow \infty$ and hence $\Omega_{\mathbf{k}\mathbf{k}^\prime}- \Omega_{\mathbf{k}^\prime \mathbf{k}}\rightarrow 0$. Therefore, a finite magnetic field is essential to the skew scattering mechanism, which breaks the time-reversal symmetry.

\subsection{Coordinate jump under magnetic field}

In this section, we discuss the coordinate jump~\cite{PhysRevB.2.4559, PhysRevB.72.045346, PhysRevB.73.075318}, labeled by $\delta\mathbf{r}_{\mathbf{k}^\prime \mathbf{k}}$ (coordinate jump from $\mathbf{k} \rightarrow \mathbf{k'}$). In our recipe of describing the impurity scattering, it can be conveniently defined as the difference between the starting point $\mathbf{r}_s$ and the ending point $\mathbf{r}_e$: $\delta\mathbf{r}_{\mathbf{k}^\prime \mathbf{k}}=\mathbf{r}_e-\mathbf{r}_s$. It can be further divided into longitudinal jump and transverse jump, which are parallel and orthogonal to the incoming momentum $\mathbf{k}$, respectively (Fig.~\ref{fig_traditional hard ball}). 

As the incoming momentum is along $x$-axis, the longitudinal jump is $\delta\mathbf{x}_{\mathbf{k}^\prime \mathbf{k}}$, and the transverse jump is $\delta\mathbf{y}_{\mathbf{k}^\prime \mathbf{k}}$. Similar to the differential cross section, the coordinate jump is also a functions of $\theta$ and $k$, and can be calculated as follows based on the two-dimensional Lorentz model (derived in Appendix \ref{APP-B})
\begin{align} \label{eq_longj}
\delta\mathbf{x}_{\mathbf{k}^\prime \mathbf{k}}&=R\left[  \sin\theta
-\frac{\sin\theta+2\frac{a}{R}\sin\left(  \frac{\theta}{2}\right)  }%
{\sqrt{1+\frac{2a}{R}\cos(\frac{\theta}{2})+\frac{a^{2}}{R^{2}}}}\right]\mathbf{\hat{x}}\,,\\
\label{eq_tranj}\delta\mathbf{y}_{\mathbf{k}^\prime \mathbf{k}}&=2R\sin^{2}\left(
\frac{\theta}{2}\right)  \left[  1-\frac{1}{\sqrt{1+\frac{2a}{R}\cos
(\frac{\theta}{2})+\frac{a^{2}}{R^{2}}}}\right] \mathbf{\hat{y}}\,.
\end{align}

Generally, the coordinate jump has two contributions to the electron transport. First, it may induce a net jump velocity $\mathbf{v}_{cj}$ that modifies the electronic drift velocity:
\begin{equation}
\mathbf{v}_{cj}=\sum_{\mathbf{k}^\prime} W_{ \mathbf{k} \mathbf{k}^\prime} \delta \mathbf{r}_{\mathbf{k}^\prime \mathbf{k}}=\int_{0}^{2\pi}d\theta n_{i}v\Omega_{\mathbf{k}\mathbf{k}^{\prime}}\delta\mathbf{r}_{\mathbf{k}^\prime \mathbf{k}}, 
\label{vcj}
\end{equation}
with $v=\hbar k/m$. 
Secondly, it leads to an electrostatic potential difference $e\mathbf{E}\cdot \delta \mathbf{r}_{\mathbf{k}^\prime \mathbf{k}}$ and thus affects the electronic equilibrium distribution function. 

Finally, we comment that as $\mathbf{B}\rightarrow 0$ the transverse jump does not have a net jump velocity, as the system respects a mirror symmetry with the mirror passing through the scatterer, parallel to $\mathbf{k}$, and normal to the material plane. On the other hand, the longitudinal jump is not restricted by any symmetry and hence the net jump velocity is nonzero. Both statements can be easily verified for the two-dimensional Lorentz model using Eq.~\ref{eq_kkprime}, \ref{eq_longj}, \ref{eq_tranj} and  \ref{vcj}.

\subsection{The nature of the anisotropic scattering}

In the first glance, the assignment of the scattering events of $t=0$ at the event line instead of the circular boundary of scatter is counterintuitive and artificial. However, it has deeper physical ground underneath. 

\begin{figure}[h]
\setlength{\abovecaptionskip}{0pt}
\setlength{\belowcaptionskip}{0pt}
\scalebox{0.19}{\includegraphics*{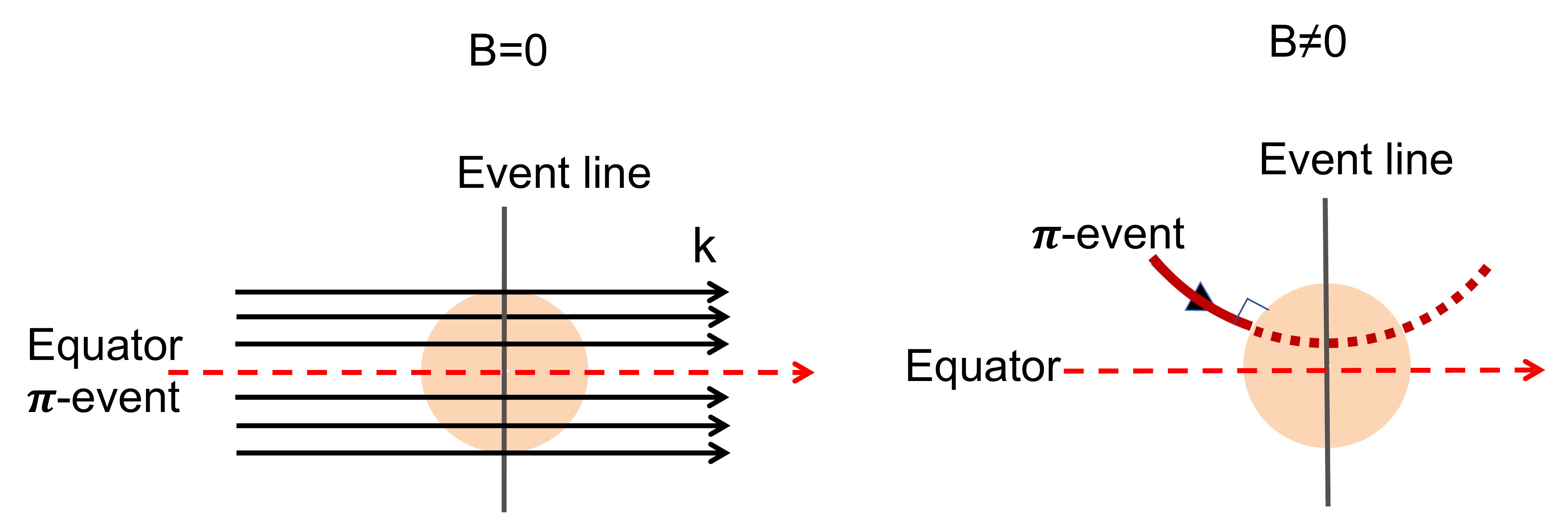}}
\caption{The illustration of `$\pi$ event' (when the scattering angle is $\pi$) in the absence and presence of magnetic field. }
\label{fig_pi_not_0}
\end{figure}

The advantage of using the event line defined in our theory instead of the colliding boundary, is that the cross sectional area (which overlaps with the event line) is the projection of the boundary. The incoming scattering events are uniformly distributed on the event line with momentum perpendicular to the event line, but not uniform on the boundary. Therefore, the number of electrons being scattered is proportional to the cross-sectional area on the event line. This provides convenience to count the number of scattering events and scattering cross section. 

\begin{figure}[h]
\setlength{\abovecaptionskip}{0pt}
\setlength{\belowcaptionskip}{0pt}
\scalebox{0.35}{\includegraphics*{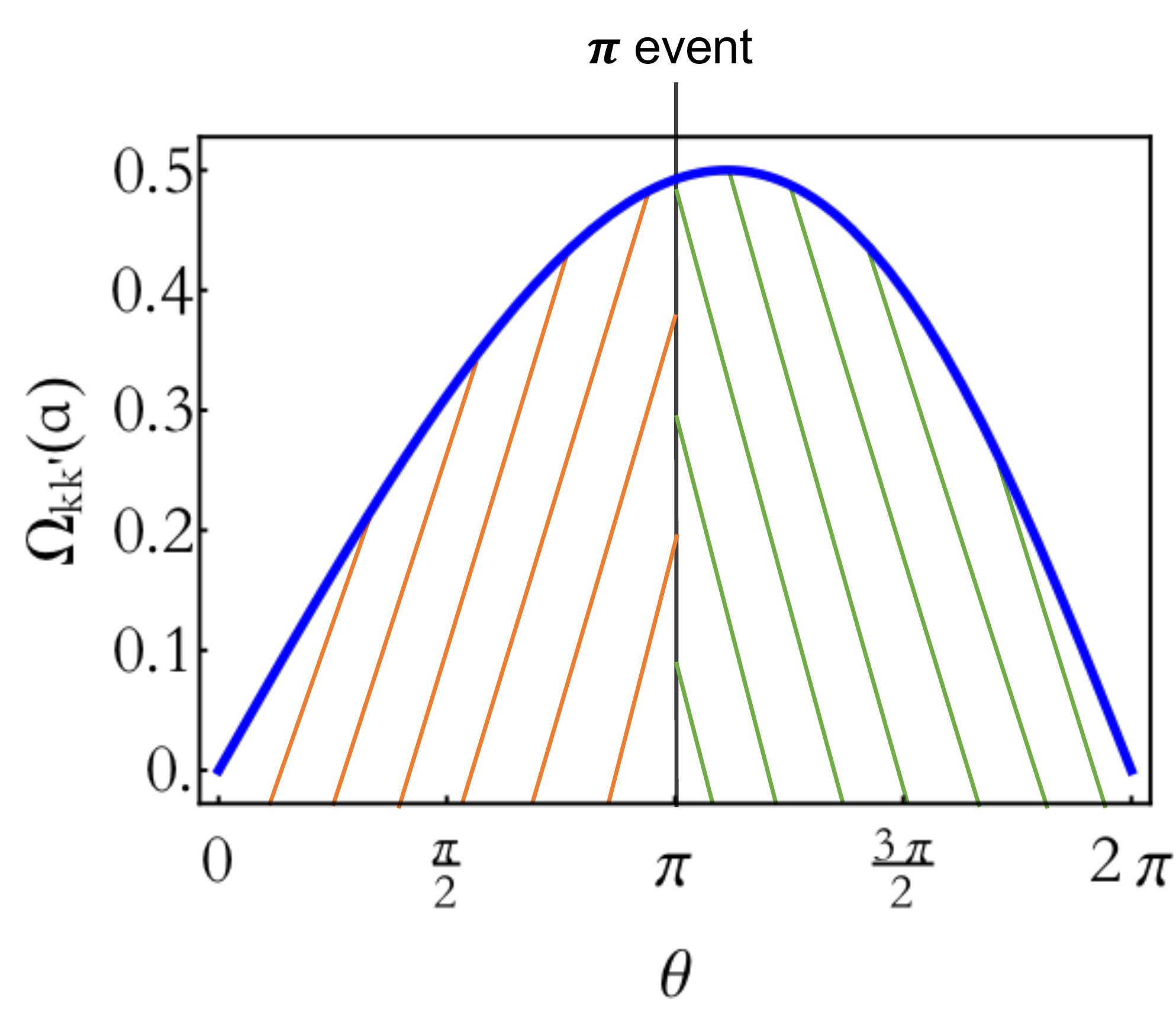}}
\caption{The plot of differential cross section of $\mathbf{k} \rightarrow \mathbf{k}^{\prime}$ process in the unit of impurity radius $a$. The vertical black line marks the $\pi$ event. The red shaded area is the cross sectional area within scattering angle $[0,\pi]$. The green shaded area is the cross sectional area within scattering angle $[\pi,2\pi]$. The $\pi$-event unevenly divides the cross sectional area, with the red shaded area smaller than the green shaded area. }
\label{fig_cross_theta}
\end{figure}

In order to understand the nature of anisotropic scattering, we define `$\pi$ event' as the scattering event with scattering angle $\theta=\pi$. When there is no magnetic field, the `$\pi$ event' evenly divides the cross sectional area along the event line (Fig. \ref{fig_pi_not_0}) and there is no skew scattering. When there is magnetic field present, the $\pi$ event unevenly divides the cross-sectional area on the event line (Fig. \ref{fig_pi_not_0}), resulting in the uneven division of the number of electrons being scattered up (with the scattering angle within $[0, \pi]$) and scattered down (with the scattering angle within $[\pi, 2\pi]$). This is shown in Fig. \ref{fig_cross_theta}, where the red shaded area (corresponding to the cross-sectional area being scattered up) is smaller than the green shaded area (corresponding to the cross-sectional area being scattered down). 

We provide a second way to understand the anisotropic scattering in Appendix \ref{APP-F}.

\subsection{Modified Boltzmann equation}

The Boltzmann equation can be generalized to include the skew scattering and coordinate jump, reading as ($e>0$)%
\begin{widetext}
\begin{equation}
\left(  -e\right)  \left(  \mathbf{E+v}\times\mathbf{B}\right)  \cdot
\frac{\partial f_{\mathbf{k}}}{\hbar\partial\mathbf{k}}=-n_{i}v\int_{0}^{2\pi
}d\theta \left [ \Omega_{\mathbf{kk}^{\prime}}  f\left(  \epsilon
,\mathbf{k}\right)  -\Omega_{\mathbf{k}^{\prime}\mathbf{k}} f\left(  \epsilon,\mathbf{k}^{\prime}\right)
+\Omega_{\mathbf{k}^\prime \mathbf{k}}\partial_{\epsilon}f^{0}e\mathbf{E}\cdot\delta\mathbf{r}_{\mathbf{k}^{\prime
}\mathbf{k}}\right ]  ,
\label{Boltzmann}
\end{equation}
\end{widetext}
where $f^0$ is the equilibrium distribution function. We emphasize that in the above equation, $|\mathbf{k}^\prime|=|\mathbf{k}|$ because the scattering is elastic.

To solve up to the linear order of electric field, we assume that
\begin{equation}
f\left(  \epsilon,\mathbf{k}\right)  =f^{0}\left(  \epsilon\right)
+g^{\rm r}\left(  \epsilon,\mathbf{k}\right)  +g^{\rm cj}\left(  \epsilon
,\mathbf{k}\right)  ,
\label{assume}
\end{equation}
where $g^{\rm cj}\left(  \epsilon,\mathbf{k}\right)$ is the part of the non-equilibrium distribution function purely due to the coordinate jump (or called anomalous distribution function), and $g^{\rm r}$ is the non-equilibrium distribution function in the absence of coordinate jump (or called normal distribution function).
Combining Eq.~\ref{Boltzmann} and Eq.~\ref{assume}, keeping the terms of linear order in the electric and magnetic field, and ignoring the coupling between skew scattering and coordinate jump, the Boltzmann equation is decomposed into two equations:%
\begin{widetext}
\begin{equation}
\left(  -e\right)  \mathbf{E}\cdot\frac{\partial f^{0}}{\hbar\partial
\mathbf{k}}+\left(  -e\right)  \left(  \mathbf{v}\times\mathbf{B}\right)
\cdot\frac{\partial g^{\rm r}_{\mathbf{k}}}{\hbar\partial\mathbf{k}}=-\int_{0}^{2\pi
}d\theta n_{i}v \left[ \Omega_{\mathbf{kk}^{\prime}} g^{\rm r}_{\mathbf{k}%
}-\Omega_{\mathbf{k}^{\prime}\mathbf{k}} g^{\rm r}_{\mathbf{k}\prime}\right]  ,\label{SBE-n}%
\end{equation}%
\begin{equation}
\left(  -e\right)  \mathbf{E}\cdot\left(  \int_{0}^{2\pi}d\theta n_{i}%
v\Omega_{\mathbf{k}^{\prime} \mathbf{k}}\delta\mathbf{r}_{\mathbf{k}^{\prime}\mathbf{k}}\right)  \partial_{\epsilon}f^{0}-\left(  -e\right)  \left(
\mathbf{v}\times\mathbf{B}\right)  \cdot\frac{\partial g_{\mathbf{k}}^{\rm cj}%
}{\hbar\partial\mathbf{k}}=\int_{0}^{2\pi}d\theta n_{i}v \left[ \Omega_{\mathbf{kk}%
^{\prime}}  g_{\mathbf{k}}^{\rm cj}-\Omega_{\mathbf{k}^{\prime}\mathbf{k}} g_{\mathbf{k}\prime}^{\rm cj}\right].
\label{SBE-a}%
\end{equation}
\end{widetext}

With all the above ingredients
the electrical current density is given by%
\begin{equation}
\mathbf{j}=\left(  -e\right)  \int \frac{d\mathbf{k}}{4\pi^2}\left[  g^{\rm r}+g^{\rm cj}\right]  \left[
\mathbf{v}+\mathbf{v}^{cj}\right]  .
\end{equation}

\section{Solutions of the Boltzmann equation}

\subsection{Zero magnetic field case}

In this case, only the longitudinal coordinate jump along the $\mathbf{k}$-direction exists. 
$\mathbf{v}^{cj}\equiv\int_{0}^{2\pi}d\theta n_{i}v\Omega
\left(  \theta\right)  \delta\mathbf{r}_{\mathbf{k}^{\prime}\mathbf{k}}=-\mathbf{v}\frac{3\pi n_{i}a^{2}}{4}$ which is along the opposite direction to
$\mathbf{v}$. 

The Boltzmann equation is solved as%
\begin{equation}
g^{\rm r}_{\mathbf{k}}=\left(  -\partial_{\epsilon}f^{0}\right)  \left(  -e\right)
\mathbf{E\cdot v}\tau^{0}\left(  \epsilon\right)  ,
\end{equation}
\begin{equation}
g_{\mathbf{k}}^{\rm cj}=\left(  \partial_{\epsilon}f^{0}\right)  \left(
-e\right)  \mathbf{E\cdot v}^{cj}\tau^{0}\left(  \epsilon\right)  ,
\end{equation}
where $\frac{1}{\tau^{0}\left(  \epsilon\right)  }=n_{i}v\frac{8a}{3}%
$. The electric current density is therefore
$j_{x}\equiv\left(  \sigma^{0}+\sigma^{cj1}+\sigma^{cj2}+\sigma^{cj1,cj2}\right)
E_{x}$ with%
\begin{align}
\sigma^{0}=\left(  -e\right)  \sum_{k}\frac{g^{\rm r}_{\mathbf{k}}}{E_{x}}v_{x}%
=\frac{ne^{2}\tau^{0}\left(  \epsilon_{F}\right)  }{m},
\end{align}
\begin{align}
\sigma^{\rm cj1} &  =\left(  -e\right)  \sum_{k}\frac{g_{\mathbf{k}}^{\rm cj}}{E_{x}%
}v_{x}=\frac{3n_{i}\pi a^{2}}{4}\frac{ne^{2}\tau^{0}\left(
\epsilon_{F}\right)  }{m},\\
\sigma^{\rm cj2} &  =\left(  -e\right)  \sum_{k}\frac{g^{\rm r}_{\mathbf{k}}}{E_{x}}%
v_{x}^{\rm cj}=-\sigma^{\rm cj1},
\label{cancel}
\end{align}
and%
\begin{align}
\sigma^{\rm cj1, \rm cj2}=\left(  -e\right)  \sum_{k}\frac{g_{\mathbf{k}}^{\rm cj}}{E_x} v_{x}^{\rm cj}%
=-\frac{ne^{2}\tau^{0}\left(  \epsilon_{F}\right)  }{m}\left(
\frac{3n_{i}\pi a^{2}}{4}\right)  ^{2},
\end{align}
where the carrier density $n=\frac{m \epsilon_{F}}{\pi\hbar^{2}}$ with $\epsilon_{F}$ the Fermi energy.\ 

Here, $\sigma^{0}$ is the conventional zero-field conductivity in the Drude theory. $\sigma^{\rm cj1}$ is the conductivity induced by the anomalous distribution from the coordinate jump. $\sigma^{\rm cj2}$ is the conductivity induced by the velocity correction from the coordinate jump. It cancels $\sigma^{\rm cj1}$. $\sigma^{\rm cj1,\rm cj2}$ is the conductivity with both the distribution and velocity being corrected by the coordinate jump. 
Therefore, the total electrical conductivity is
\begin{equation}
\sigma=\sigma^{0}+\sigma^{\rm cj1, \rm cj2}=\frac{ne^{2}\tau^{0}\left(
\epsilon_{F}\right)  }{m}\left[  1-\left(  \frac{3}{4}n_{i}\pi a^{2}\right)
^{2}\right]  .
\end{equation}

There is a correction to the electron density, because the electrons are only present in the
free area excluding the area occupied by impurities. The electron density
$n=\frac{N}{A-A_i}=\frac{n_D}{1-\frac{A_i}{A}}$,
where $A$ and $A_{i}$ represent the total 2D area and
the area occupied by the hard disk impurities, respectively, and $\frac{A_i}{A}=\pi n_i a^2$, and
$n_D=\frac{N}{A}$ is the electron density without the correction to exclude the area that impurities take. 
Thus, the Fermi momentum $k_F=\sqrt{2\pi n}=\frac{{k_F}^D}{\sqrt{1-\frac{A_i}{A}}}$, where $k_{F}^{D}=\sqrt{2\pi n_{D}}$. 

Therefore, the measured electrical conductivity is also corrected by
\begin{equation}
\begin{aligned}
\sigma^{M}&=\sigma\frac{A-A_{i}}{A}\\
                  &=\frac{n_{D}e^{2}\tau^{D} }{m}\left[  1-\left(  \frac{3}{4}\pi n_{i}a^{2}\right)
^{2}\right]  \sqrt{1-\pi n_{i}a^{2}},
\label{corrected cond}
\end{aligned}
\end{equation}
with the Drude transport relaxation rate $1/\tau^{D}=n_{i}v_{F}^{D}\frac
{8a}{3}$ a constant. 
The conductivity in our theory $\sigma^{M}$ is lower than the Drude conductivity $\sigma^{D}=\frac{n_{D}e^{2}\tau^{D} }{m}$ by a factor of $\left[  1-\left(  \frac{3}{4}\pi n_{i}a^{2}\right)^{2}\right]  \sqrt{1-\pi n_{i}a^{2}}$ as can be seen from Eq.~\ref{corrected cond}, which decreases as a function of the dimensionless quantity $n_{i}a^{2}$. The deviation of the diffusion coefficient from the Drude model in a previous computer simulation of Lorentz model with overlapped hard sphere impurities \cite{PhysRevA.25.533} is similar to that in our theory. 

\subsection{Low magnetic field case: Hall coefficient and magnetoresistivity}

In this section, we evaluate the conductivity under a weak magnetic field. We first discuss the contribution from the skew scattering. According to previous discussions, we need to solve the distribution function using Eq. ~\ref{SBE-n}. 

$g_{\mathbf{k}}^{\rm r}=\left(  -\partial_{\epsilon}f^{0}\right)  \left(  -e\right)
\left[  \mathbf{E\cdot v}\tau^{L}\left(  \epsilon\right)  +\left(
\mathbf{\mathbf{\hat{z}}\times E}\right)  \cdot\mathbf{v}\tau^{T}\left(
\epsilon\right)  \right]$ into Eq.~\ref{SBE-n} and obtain%
\begin{align}
\tau^{L}\left(  \epsilon\right)   &  =\frac{\tau^{\parallel}\left(
\epsilon\right)  }{1+\left[  \omega_{c}\tau^{\parallel}\left(  \epsilon
\right)  +\frac{\tau^{\parallel}\left(  \epsilon\right)  }{\tau^{\perp}\left(
\epsilon\right)  }\right]  ^{2}},\nonumber\\
\tau^{T}\left(  \epsilon\right)   &  =\left[  \omega_{c}\tau^{\parallel
}\left(  \epsilon\right)  +\frac{\tau^{\parallel}\left(  \epsilon\right)}{\tau^{\perp}\left(  \epsilon\right)  }\right]  \tau^{L}\left(\epsilon\right),
\end{align}
where we define%
\begin{widetext}
\begin{align}
\frac{1}{\tau^{\parallel}\left(  \epsilon\right)  }  &  =\int_{0}^{2\pi
}d\theta n_{i}v [ \Omega^{A}\left(  1+\cos\left(
\theta \right)  \right)+\Omega^{S}\left( 1-\cos\left(\theta \right)  \right) ] =\frac{8}{3}n_{i}va\left[  1-\frac{1}{5}\left(  \frac{a}{R}\right)  ^{2}+O\left(  \left(  \frac{a}{R}\right)  ^{4}\right) \right],
\label{tau-para}
\end{align}
\begin{align}
\frac{1}{\tau^{\perp}\left(  \epsilon\right)  }  &  =\int_{0}^{2\pi}d\theta
n_{i}v [ \Omega^{S}-\Omega^{A} ] \sin\left(  \theta \right)  =-\frac{\pi}{4}n_{i}va\frac{a}{R}\left[  1+O\left(  \left(  \frac{a}{R}\right)  ^{2}\right)
\right]. 
\label{tau-perp}
\end{align}
\end{widetext}

Here $\Omega^{A}=\frac{1}{2}\left(\Omega_{\mathbf{kk}^{\prime}}-\Omega_{\mathbf{k}^{\prime}\mathbf{k}}\right)  $, which is the antisymmetric part of the differential cross section, and $\Omega^{S}=\frac{1}{2}\left(\Omega_{\mathbf{k}^{\prime}\mathbf{k}}+\Omega_{\mathbf{kk}^{\prime}}\right)  $, which is the symmetric part of the differential cross section. $\tau^{\perp}$ is purely due to the skew scattering, i.e. $\Omega_\mathbf{kk'} \neq \Omega_\mathbf{k'k}$. In our theory, only when $B \neq 0$, $\Omega_\mathbf{kk'} \neq \Omega_\mathbf{k'k}$. 

Generally, we prove that $\tau^{\parallel}$ is purely contributed  by $\Omega^{S}$ by showing $\int_{0}^{2\pi}d\theta \Omega^{A} (1+\cos (  \theta ))=0$, and $\tau^{\perp}$ is purely contributed by $\Omega^{A}$ by showing  $\int_{0}^{2\pi}d\theta \Omega^{S} \sin\left(  \theta \right)=0$ (see Appendix \ref{APP-D}). 
As a result, $\tau^{\parallel}$ is not enough to characterize the collision process as long as the scattering probability contains an antisymmetric part, in which case, $\tau^{\perp}$ naturally emerges. 

\begin{figure}[h]
\setlength{\abovecaptionskip}{0pt}
\setlength{\belowcaptionskip}{0pt}
\scalebox{0.36}{\includegraphics*{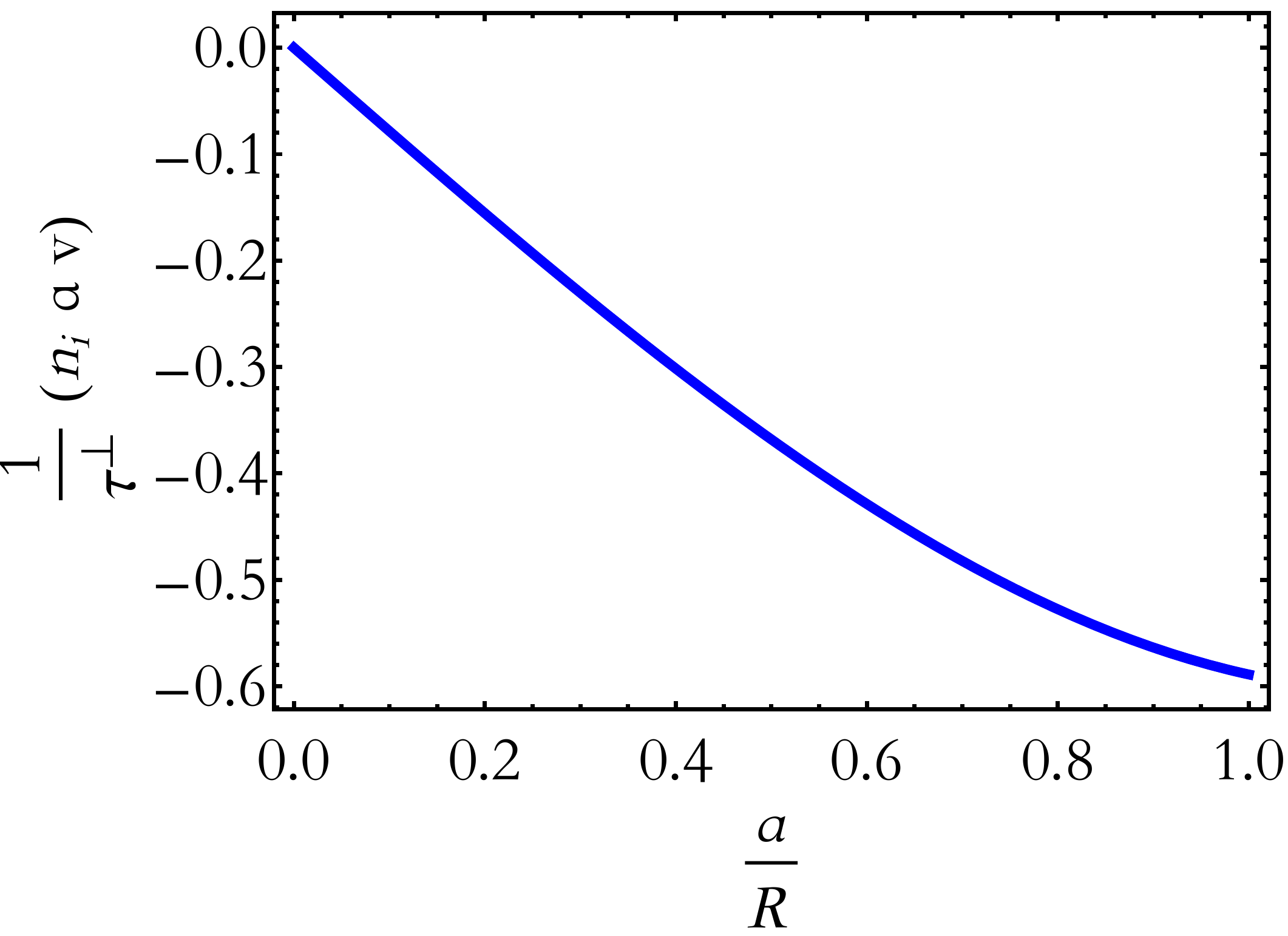}}
\caption{The plot of the reciprocal of transverse relaxation time $\frac{1}{\tau^{\perp}}$ in the unit of $n_{i} a v$, where $a$ is the impurity radius, $n_{i}$ is the impurity density, and  $v$ is the electron velocity. The $\frac{1}{\tau^{\perp}}$ is always negative as long as $a<R$. }
\label{fig_tau_perp}
\end{figure}

In our example, $\frac{1}{\tau^{\perp}}$ is always negative (as shown in Eq. \ref{tau-perp} and Fig. \ref{fig_tau_perp}). Besides, as Eq. \ref{tau-perp} shows, $\frac{1}{\tau^{\perp}} \neq 0$, as long as $\frac{a}{R}$ is finite. Moreover, the ratio of $\tau^{\parallel}$  to $\tau^{\perp}$ is proportional to $a/R$. Since we are considering the weak magnetic field scenario with a large $R$ ($R>a$), $\tau^{\perp}$ will be bigger than $\tau^{\parallel}$. Taking the data from the second row of Table I as example where $\beta=0.6$, $\frac{a}{R}=\frac{ 2\beta  c}{\pi} =0.06$, the ratio of $\tau^{\parallel}$  to $\tau^{\perp}$ is then around $-0.017$.

The conductivity resulted from the skew scattering is 
\begin{equation}
\left[
\begin{array}
[c]{c}%
\sigma_{xx}\\
\sigma_{yx}%
\end{array}
\right]  =\frac{\frac{ne^{2}\tau^{\parallel}}{m}}{1+\left(  \omega_{c}\tau^{\parallel}+\frac{\tau^{\parallel}%
}{\tau^{\perp}}\right)  ^{2}}\left[
\begin{array}
[c]{c}%
1 \\
(eB+\frac{m}{\tau^{\perp}}) \frac{\tau^{\parallel}}{m}
\end{array}
\right]  .\label{con-skew}%
\end{equation}

Converting the conductivity in Eq. \ref{con-skew} into resistivity, we get 
\begin{eqnarray}
\rho_{xx}=\frac{m}{e^{2} \tau^{\parallel} n}, 
\label{rouxx}
\end{eqnarray}
\begin{eqnarray}
\rho_{xy}=-\left( \frac{B}{en}+\frac{m}{e^2 \tau^{\perp} n} \right). 
\label{rouxy}
\end{eqnarray}
where $\frac{1}{\tau^{\parallel}}$ and $\frac{1}{\tau^{\perp}}$ was derived in Eq. \ref{tau-para} and Eq. \ref{tau-perp}, respectively. 

As Eq. \ref{rouxx} and Eq. \ref{rouxy} shows, $\tau^{\parallel}$ contributes to the negative magnetoresistance, while $\tau^{\perp}$ contributes to the anomalous Hall effect. 
The $\frac{1}{\tau^{\parallel}}$ decreases with the increase of magnetic field, which results in the negative magnetoresistance. We will explore more of $\tau^{\perp}$ from Drude theory perspective in Section IV Discussion, part C. 

We now discuss the contribution from coordinate jump to the conductivity. For the anomalous distribution function due to coordinate jump, we first calculate%
\begin{eqnarray}
\int_{0}^{2\pi}d\theta n_{i}v\Omega_{\mathbf{k}^{\prime}\mathbf{k}}\delta
\mathbf{r}_{\mathbf{k}^{\prime}\mathbf{k}} =n_{i}a^{2}\left(
C_{a}^{\parallel}\mathbf{v}+C_{a}^{\perp}\mathbf{\hat{z}}\times\mathbf{v}\right),
\end{eqnarray}
and 
\begin{eqnarray}
\int_{0}^{2\pi}d\theta n_{i}v\Omega_{\mathbf{kk}^{\prime}}\delta
\mathbf{r}_{\mathbf{k}^{\prime}\mathbf{k}} =n_{i}a^{2}\left(
C_{cj}^{\parallel}\mathbf{v}+C_{cj}^{\perp}\mathbf{\hat{z}}\times\mathbf{v}\right),
\end{eqnarray}
where %
\begin{eqnarray}
C_{a}^{\parallel} &  \equiv-\frac{3}{4}\pi+\frac{7\pi}{16}\left(  \frac{a}
{R}\right)  ^{2}+O\left(  \frac{a}{R}\right)  ^{4},
\end{eqnarray}
\begin{eqnarray}
C_{a}^{\perp}\equiv\frac{16}{15}\frac{a}{R}+\frac{88}{105}\left(  \frac{a}{R}\right)
^{3}+O\left(  \frac{a}{R}\right)  ^{5},
\end{eqnarray}
\begin{eqnarray}
C_{cj}^{\parallel} &  \equiv-\frac{3}{4}\pi+\frac{\pi}{16}\left(  \frac{a}
{R}\right)  ^{2}+O\left(  \frac{a}{R}\right)  ^{4},
\end{eqnarray}
\begin{eqnarray}
C_{cj}^{\perp}\equiv\frac{8}{105}\left(  \frac{a}{R}\right)  ^{3}+O\left(  \frac{a}
{R}\right)  ^{5} .
\end{eqnarray}

Then we plug the following ansartz%
\begin{equation}
g_{\mathbf{k}}^{\rm cj}=\left(  -\partial_{\epsilon}f^{0}\right)  \left(
-e\right)  n_{i}a^{2}\left[  \mathbf{E\cdot v}\tau^{L,\rm cj}\left(  \epsilon
\right)  +\left(  \mathbf{\hat{z}\times E}\right)  \cdot\mathbf{v}%
\tau^{T,\rm cj}\left(  \epsilon\right)  \right],  
\end{equation}
into Eq. ~\ref{SBE-a} and get%
\begin{align}
\tau^{L,cj}\left(  \epsilon\right)   &  =\frac{-C_{a}^{\parallel}-C_{a}^{\perp
}\left[  \omega_{c}\tau^{\parallel}\left(  \epsilon\right)  +\frac
{\tau^{\parallel}\left(  \epsilon\right)  }{\tau^{\perp}\left(  \epsilon
\right)  }\right]  }{1+\left(  \omega_{c}\tau^{\parallel}\left(
\epsilon\right)  +\frac{\tau^{\parallel}\left(  \epsilon\right)  }{\tau
^{\perp}\left(  \epsilon\right)  }\right)  ^{2}}\tau^{\parallel}\left(
\epsilon\right)  ,\\
\tau^{T,cj}\left(  \epsilon\right)   &  =\frac{C_{a}^{\perp}-C_{a}^{\parallel
}\left(  \omega_{c}\tau^{\parallel}\left(  \epsilon\right)  +\frac
{\tau^{\parallel}\left(  \epsilon\right)  }{\tau^{\perp}\left(  \epsilon
\right)  }\right)  }{1+\left(  \omega_{c}\tau^{\parallel}\left(
\epsilon\right)  +\frac{\tau^{\parallel}\left(  \epsilon\right)  }{\tau
^{\perp}\left(  \epsilon\right)  }\right)  ^{2}}\tau^{\parallel}\left(
\epsilon\right)  .
\end{align}

Combining the skew scattering and coordinate jump contributions to the conductivity, we obtain the following result
\begin{align}
\sigma_{||} &  \equiv\frac{j_{x}}{E_{x}}=\mathbf{\sigma}_{xx}\left(
1+\tan^{2}\theta_{H}\right)  ,\\
R_{H}\left(  \epsilon_{F},B \right)   &  \equiv\frac{E_{y}}{j_{x} B }=-\frac
{1}{B}\frac{\tan\theta_{H}}{\mathbf{\sigma}_{xx}\left(  1+\tan^{2}\theta
_{H}\right)  }, \label{RH}%
\end{align}
where $\tan\theta_{H}\equiv\frac{\mathbf{\sigma}_{yx}}{\mathbf{\sigma}_{xx}}$
denotes the Hall angle, and the Hall coefficient $R_{H}$ depends on the magnetic field and
$\epsilon_{F}$. Usually only the magnetic field-independent Hall coefficient $R_{H}|_{B=0}$
is needed. The magnetoresistivity is $\frac{\delta\rho_{xx}\left(B\right)  }%
{\rho_{xx}\left(B=0\right)  }=-\frac{\delta\mathbf{\sigma}%
_{xx}\left(B\right)  }{\mathbf{\sigma}_{xx}\left(B=0\right)
}$ where $\rho_{xx}=\mathbf{\sigma}_{xx}^{-1}$ is the
resistivity and $\delta\mathbf{\sigma}_{xx}\left(B\right)
\equiv\mathbf{\sigma}_{xx}\left(B\right)  -\mathbf{\sigma}%
_{||}\left(B=0\right)  $ is the magnetoconductivity. Because%
\begin{widetext}
\begin{equation}
\left[
\begin{array}
[c]{c}%
\sigma_{xx}\\
\sigma_{yx}%
\end{array}
\right]  =\frac{e^{2}\epsilon_{F}}{\pi\hbar^{2}}\left[
\begin{array}
[c]{cc}%
1+n_{i}a^{2}C_{cj}^{\parallel} & -n_{i}a^{2}C_{cj}^{\perp}\\
n_{i}a^{2}C_{cj}^{\perp} & 1+n_{i}a^{2}C_{cj}^{\parallel}%
\end{array}
\right]  \left[
\begin{array}
[c]{c}%
\tau^{L}+n_{i}a^{2}\tau^{L,cj}\\
\tau^{T}+n_{i}a^{2}\tau^{T,cj}%
\end{array}
\right].  \label{con}%
\end{equation}
\end{widetext}

In the case of $n_{i}a^{2}\ll1$ and $\omega_{c} \tau^{0}<1$, we can neglect $C_{cj}^{\perp}$ and obtain%
\begin{widetext}
\begin{equation}
\left[
\begin{array}
[c]{c}%
\sigma_{xx}\\
\sigma_{yx}%
\end{array}
\right]  =\frac{ne^{2}}{m}\frac{\left(  1+n_{i}a^{2}C_{cj}^{\parallel}\right)
\tau^{\parallel}}{1+\left(  \omega_{c}\tau^{\parallel}+\frac{\tau^{\parallel}%
}{\tau^{\perp}}\right)  ^{2}}\left[
\begin{array}
[c]{c}%
1+n_{i}a^{2}\left[  -C_{a}^{\parallel}-C_{a}^{\perp}\left(  \omega_{c}%
\tau^{\parallel}+\frac{\tau^{\parallel}}{\tau^{\perp}}\right)  \right]  \\
\left(  \omega_{c}\tau^{\parallel}+\frac{\tau^{\parallel}}{\tau^{\perp}%
}\right)  +n_{i}a^{2}\left[  C_{a}^{\perp}-C_{a}^{\parallel}\left(  \omega
_{c}\tau^{\parallel}+\frac{\tau^{\parallel}}{\tau^{\perp}}\right)  \right]
\end{array}
\right].
\end{equation}
\end{widetext}
This is the complete expression of the conductivity including the skew scattering and coordinate jump effect. Then we can solve for the Hall coefficient and the magnetoresistivity. 

The full expression of Hall angle is given by
%\[
\begin{equation}
\tan\theta_{H}=\frac{\left(  \omega_{c}\tau^{\parallel}+\frac{\tau^{\parallel
}}{\tau^{\perp}}\right)  +n_{i}a^{2}\left[  C_{a}^{\perp}-C_{a}^{\parallel
}\left(  \omega_{c}\tau^{\parallel}+\frac{\tau^{\parallel}}{\tau^{\perp}%
}\right)  \right]  }{1+n_{i}a^{2}\left[  -C_{a}^{\parallel}-C_{a}^{\perp
}\left(  \omega_{c}\tau^{\parallel}+\frac{\tau^{\parallel}}{\tau^{\perp}%
}\right)  \right]  }.
\end{equation}
%\]

To expand the Hall coefficient and the magnetoresistivity, we use the approximation $n_{i}a^{2}\ll1$.
The Hall angle and the magnetoconductivity are%
\begin{equation}
\tan\theta_{H}\simeq\omega_{c}\tau^{0}\left[  1-\frac{\pi}{4}%
n_{i}a^{2}+\frac{128}{45}\left(  n_{i}a^{2}\right)  ^{2}\right]  ,
\label{Hall angle}%
\end{equation}
and
\begin{equation}
\frac{\sigma_{||}\left(B \right)  -\sigma_{||}\left(  0\right)  }{\sigma
_{||}\left(  0\right)  }\simeq\frac{64}{15}\left(  n_{i}a^{2}\omega_{c}%
\tau^{0}\right)  ^{2}=\frac{3}{5}\left(  \frac{a}{R}\right)  ^{2}>0,
\end{equation}
respectively. The correction due to the effective area of free space excluding the area of all the impurities is of higher order and thus neglected.

The magnetoresistivity $\frac{\delta\rho_{\parallel}\left(
B\right)  }{\rho_{\parallel}\left(  0\right)  }\simeq-\frac{64}{15}\left(
n_{i}a^{2}\omega_{c}\tau^{0}\right)  ^{2}$ is negative, and
is composed of three contributions: 1) the
contribution from the Hall angle, more specifically, from the anomalous distribution function to the Hall transport $\left(  C_{a}^{\perp
}\rightarrow\tau^{T,cj}\rightarrow\tan\theta_{H}\right)  $;
2) the magnetic-field-induced correction to the longitudinal transport
relaxation time $\left(  \left(  \tau^{\parallel}-\tau^{0}\right)
\rightarrow\sigma_{xx}\right)  $;
3) the contribution of
anomalous distribution function to the longitudinal transport $\left(
C_{a}^{\perp}\rightarrow\tau^{L,cj}\rightarrow\sigma_{xx}\right)  $.

The leading order correction of the Hall angle $-\frac{\pi}{4}n_{i}a^{2}\omega_{c}\tau^{0}$
stems from the magnetic-field-induced skew
scattering. This result is comparable to that corrected by the classical memory
effect \cite{PRB2008} in the limit $n_{i}a^{2}\ll\omega_{c}\tau_{D}%
\ll1$: $\delta R_{H}^{cm}/R_{H}^{B}=-\frac{32}{9\pi}n_{i}a^{2}$, where $R_{H}^{B}$ is the Hall coefficient in the conventional Boltzmann theory $R_{H}^{B}=-\frac{1}{n_D e}=-\frac{1}{ne(1-\pi n_i a^2)}$, and $\delta R_{H}^{cm}$ 
is the difference between the Hall coefficient corrected by classical memory effect and the conventional Hall coefficient. 

In experiments, to obtain the real electron density $n$ from the measured Hall coefficient, the correction to $R_{H}$ has to be included. 
The Hall coefficient is $R_{H}=-1/n^{\prime}e$, where
$n^{\prime}$ is the effective electron density $n^{\prime}\approx \frac{n}{\left(
1-\frac{c}{4}+\frac{128 c^{2}}{45\pi^{2}}\right)}$ and $c=\pi a^{2}n_{i}$. We use the 
value of $c=0.15$ here as an example (this value is also used in the discussion) and find that $n^{\prime}\approx \frac{n}{0.97}  $ which is equivalent to a $3\%$ error. This error is larger when the impurity density increases. 

We note that in a previous work \cite{PhysRevLett. 112. 166601}, it is already recognized that there may be corrections to the Hall coefficient. However, their result is due to the magnetic-field-affected Bloch-electron drifting motion, and is proportional to the $\frac{1}{(\tau^0)^2}$ (or equivalently, $(n_i a^2)^2$). In comparison, our correction here has different origins (magnetic-field-affected electron-impurity scattering), as well as different scaling behavior (i.e. proportional to $n_i a^2$).

\section{Discussions}

\subsection{Magnetoresistivity in comparison with simulation at low magnetic field}

In this section, we compare our theoretical results with
the analytical and numerical results for the pure 2D Lorentz model previously
obtained in literatures \cite{PRB2003, PhysRevLett.89.266804}.
At low magnetic field $\omega_{c}\tau^{0}<1$, the theory in \cite{PRB2003, PhysRevLett.89.266804} predicted a negative magnetoresistivity due to the influence of magnetic field on the Corridor effect (enhancing the backscattering from the first impurity to the second impurity and back to the first impurity) and on multiple scatterings. 

%\begin{widetext}
\begin{table*}[t]
\centering
\begin{tabular}{|c|c|c|c|c|c|c|}
\hline
$\beta$ & $\left(  \frac{\delta\rho_{\parallel}}{c\rho_{0}}\right)  ^{an}$ &
$\frac{\delta\rho_{\parallel}^{\prime}}{c\rho_{0}}$ & $\left(  \frac{\delta
\rho_{\parallel}^{Cor}}{c\rho_{0}}\right)  ^{th}$ & $\frac{\delta\rho
_{\parallel}^{\prime}}{c\rho_{0}}+\left(  \frac{\delta\rho_{\parallel}^{Cor}%
}{c\rho_{0}}\right)  ^{th}$ & $\left(  \frac{\delta\rho_{\parallel}}{c\rho_{0}%
}\right)  ^{an}+\frac{\delta\rho_{\parallel}^{\prime}}{c\rho_{0}}+\left(
\frac{\delta\rho_{\parallel}^{Cor}}{c\rho_{0}}\right)  ^{th}$ & $\left(
\frac{\delta\rho_{\parallel}}{c\rho_{0}}\right)  ^{si}$\\
\hline
$\beta/c=3,\beta=0.45$ & -0.0074 & -0.026 & -0.0605 & -0.0865 & -0.094 & -0.1\\
\hline
$\beta/c=4,\beta=0.6$ & -0.013 & -0.046 & -0.07 & -0.116 & -0.13 & -0.14\\
\hline
$\beta/c=5,\beta=0.75$ & -0.021 & -0.072 & -0.076 & -0.148 & -0.17 & -0.19\\
\hline
$\beta/c=5.5,\beta=0.825$ & -0.025 & -0.087 & -0.0796 & -0.167 & -0.19 & -0.24\\
\hline
$\beta/c=6,\beta=0.9$ & -0.03 & -0.103 & -0.086 & -0.189 & -0.22 & -0.28\\\hline
\label{table_corridor}
\end{tabular}\\
%\begin{widetext}
\caption{The comparison between the summation of analytical results from \cite{PRB2003} and our theory, and the numerical results from \cite{PhysRevLett.89.266804}. The second column $\left(  \frac{\delta\rho_{\parallel}}%
{c \rho_{0}}\right)  ^{an}$ is the magnetoresistivity calculated by our formula. The third column $\frac{\delta\rho_{\parallel}^{\prime}}{c\rho
_{0}}$ is the quadratic contribution due to the
influences of magnetic field on returns after multiple scatterings in \cite{PRB2003}. The fourth column $\left(  \frac{\delta\rho_{\parallel}^{Cor}}{c\rho_{0}}\right)
^{th}$ is the analytical values of magnetoresistivity influenced by Corridor effect 
in \cite{PRB2003}. The fifth column is the summation of all the analytical results from \cite{PRB2003}. The sixth column includes our results, in addition to the previous analytical results in the fifth column. The seventh column $\left(  \frac{\delta\rho_{\parallel}}{c\rho_{0}}\right)
^{si}$ is the simulation result of magnetoresistivity in \cite{PhysRevLett.89.266804}. }
%\end{widetext}
\end{table*}
%\end{widetext}

In table I, $c=\pi n_{i}a^{2}=0.15$, $\beta=\omega_{c}\tau=\frac{4}{3}\omega
_{c}\tau^{0}$, where $\tau=\left(  2 v n_{i}a\right)  ^{-1}$  is the
single-particle scattering time. The second column $\left(  \frac{\delta\rho_{\parallel}}%
{c \rho_{0}}\right)  ^{an}=\frac{12}{5\pi^{2}} c \beta^{2}$ is the magnetoresistivity up to the order
$O\left(  \left(  n_{i}a^{2}\omega_{c}\tau^{0}\right)  ^{2}\right)
$ calculated by our formula. The third column $\frac{\delta\rho_{\parallel}^{\prime}}{c\rho
_{0}}=-\frac{0.4}{\pi}\beta^{2}$ is the quadratic contribution due to
the influence of magnetic field on multiple scatterings in \cite{PRB2003}. The fourth column $\left(  \frac{\delta\rho_{\parallel}^{Cor}}{c\rho_{0}}\right)
^{th}$ is the analytical values of magnetoresistivity influenced by Corridor effect 
in \cite{PRB2003}. The fifth column is the summation of all the analytical results from \cite{PRB2003}. The sixth column includes our results in addition to the previous analytical results in the fifth column. The seventh column $\left(  \frac{\delta\rho_{\parallel}}{c\rho_{0}}\right)
^{si}$ is the simulation result of magnetoresistivity in \cite{PhysRevLett.89.266804}. 

\begin{figure}[h]
\setlength{\abovecaptionskip}{0pt}
\setlength{\belowcaptionskip}{0pt}
\scalebox{0.32}{\includegraphics*{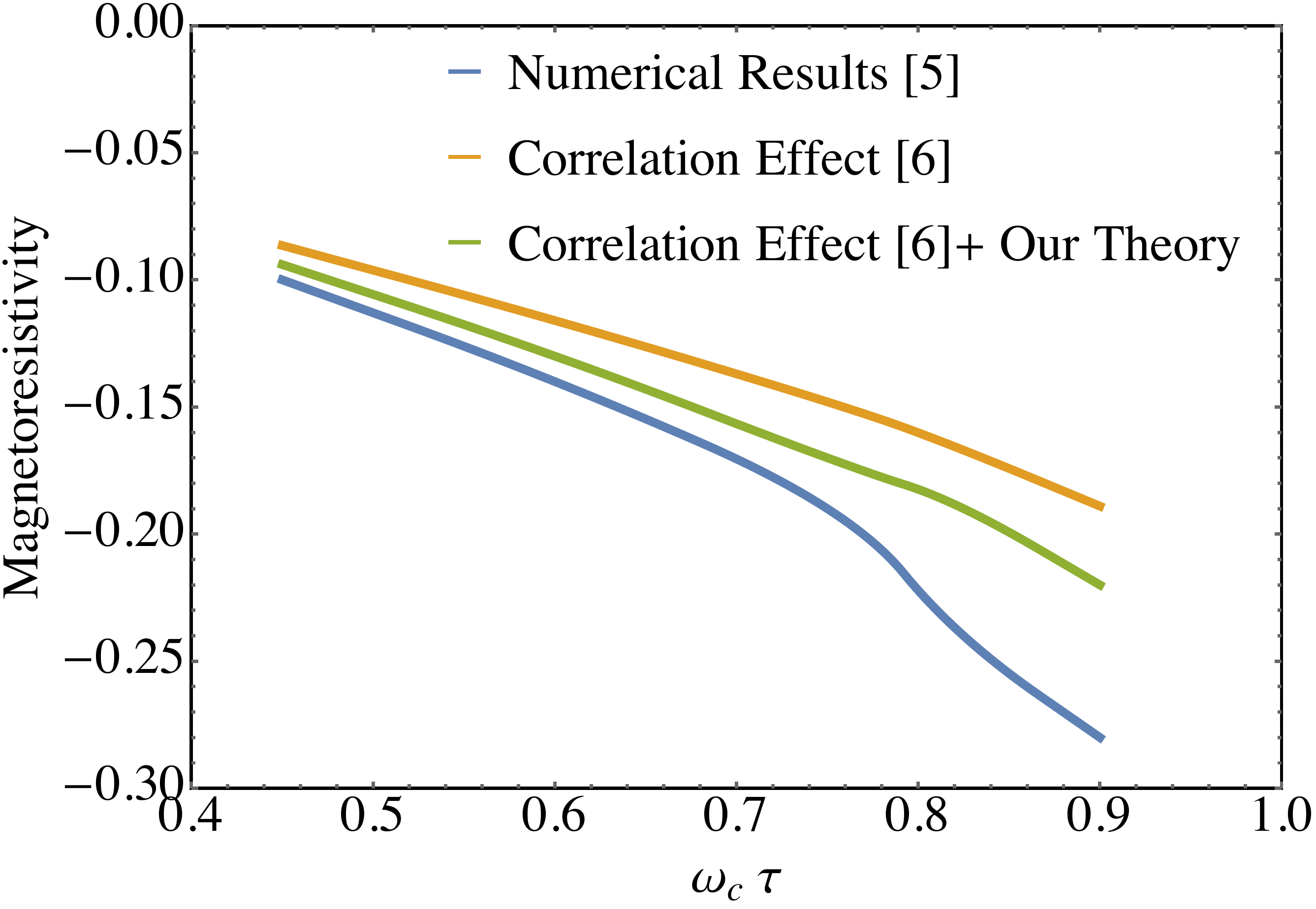}}
\caption{The plots of the numerical magnetoresistivity from [5], the magnetoresistivity from correlation effect [6], and the inclusion of our magnetoresistivity into the analytical correlation effect [6], respectively. }
\label{fig_compare}
\end{figure}

As can be seen from the table I, the
inclusion of $\left(  \frac{\delta\rho_{\parallel}}{c\rho_{0}}\right)  ^{an}%
$ (the magnetoresistivity calculated in our theory) yields a more accurate magnetoresistivity, closer to the numerical result, especially under relatively small magnetic field
($\beta=0.45$ and $\beta=0.6$). This is also reflected in Fig. \ref{fig_compare}. Under relatively larger magnetic
fields, the deviation of the analytical values from the simulation values
increases. The reason is as follows. The validity of our theory demands $R>l\gg0.6/\sqrt{n_{i}}a$ where $l$ is the mean free path \cite{PRB2001}, such that the percolation transition cannot occur.  
This requirement yields $\beta<1\ll0.83\left(  n_{i}a^{2}\right)  ^{-1/2}$.\ Using the value
$\pi n_{i}a^{2}=0.15$ in table I, we obtain the following restriction to $\beta$: $\beta \ll 3.8$. Therefore, the value of $\left(  \frac{\delta\rho_{\parallel}}{c\rho_{0}}\right)  ^{an}%
$ at large $\beta$ may not be accurate. Also,
the validity of the theory of corridor effect influenced magnetoresistivity \cite{PRB2003} holds well under similar
restrictions. Thereby the difference between the analytical (fifth column) and simulation results (sixth column) increases with larger $\beta$. 

We choose the value of $c=\pi n_{i}a^{2}=0.15$ in table I because we want to compare with the result from literatures \cite{PRB2003, PhysRevLett.89.266804} where the largest value of $c$ is $0.15$, and besides, the larger the value of $c$, the more significant the negative magnetoresistivity effect in our theory. The reason can be found from the expression $\left(  \delta\rho_{\parallel}\right)  ^{an}/\delta\rho_{\parallel}^{\prime}=6n_{i}a^{2}$. 
In literatures \cite{PRB2003, PhysRevLett.89.266804}, the dominance of corridor effect decreases when the magnetic field increases. The suppression of the corridor effect makes our result prominent, therefore, we need the magnetic field as large as possible inside the weak field regime.

\subsection{Magnetoresistivity in comparison with experimental results}

In this subsection we discuss the possible relevance of our result to
experiments. Our theory is based on the Boltzmann framework neglecting the
memory effect with successive scattering events. In real 2D electron systems with strong scatterers, the correlation between successive collisions may be
broken by the disorders and the applied electric field. Thereby, we may try to fit some experiments by only our results
regardless of the correlation effect. 

The negative parabolic magnetoresistivity has been observed in a corrugated 2DEG in GaAs wells
\cite{PRB2004}. Although the authors explained their observation by the Corridor
effect related magnetoresistivity, the fitting value for $a/l=2n_{i}a^{2}$ is beyond
its valid range (here $n_{i}a^{2}=2.6$, but the range of $n_{i}a^{2}$ is supposed to be $[0,1]$). Thus, the magnetoresistivity theory in terms of corridor effect in the 2D Lorentz model may
not provide a suitable description for the experiments with low magnetic field. If we fit the
experimental parabolic negative magnetoresistivity in low magnetic field to our formula, we
get a reasonable value $n_{i}a^{2}=0.12$. Negative parabolic magnetoresistivity was also
observed in a 2DEG in a GaN heterostructure \cite{PRB2005}, and explained by a
two-component disorder model \cite{Mirlin2001}. We can also fit the
parabolic negative magnetoresistivity by choosing $n_{i}a^{2}=0.042$. However, both of our theory and the classical magnetoresistivity theory based on memory effects \cite{PRB2005}
cannot explain the observed large negative linear magnetoresistivity in a larger magnetic field
\cite{PRB2004, PRB2005}. This experimental regime is still beyond existing theories.

We comment that in some cases, the electron motion is quasi-two-dimensional, and the vertical motion is not negligible. One example is shown in Ref. \cite{PhysRevLett.77.147}, in which an in-plane magnetic field is applied and the periodically distributed large-scale impurities are prepared. This is, however, beyond the scope of our theory.

\subsection{Phenomenological inclusion of skew scattering into the Drude model}

In this subsection we
demonstrate that the skew scattering, signified by $\frac{1}{\tau^{\perp}}$, can be phenomenologically included into Drude framework using a tensor $\tensor{\frac{1}{\tau}}$. 

In traditional Drude theory, the scattering rate $\frac{1}{\tau}$ is
treated as a scalar. The equation of motion is%
\begin{equation}
m\mathbf{\dot{v}}=-e(\mathbf{E}+\mathbf{v\times \mathbf{B}})-\frac{m \mathbf{v}}{\tau}.
\end{equation}

In the presence of out of plane magnetic field, due to the rotational symmetry in the two dimensional plane, 
$\frac{1}{\tau}$ becomes an antisymmetric tensor \cite{Physica.24.1958, PhysRevB.72.045346, Hua2015} with
nonzero off-diagonal element: 
\begin{equation}
\tensor{\frac{1}{\tau}}=\left(
\begin{array}
[c]{cc}%
\frac{1}{\tau^{\parallel}} & \frac{1}{\tau^{\perp}}\\
-\frac{1}{\tau^{\perp}} & \frac{1}{\tau^{\parallel}}%
\end{array}
\right)  .
\end{equation}

The modified equation of motion is
\begin{equation}
%\begin{aligned}
m\mathbf{\dot{v}}=-e\left(
\begin{array}
[c]{c}%
E_{x}\\
E_{y}%
\end{array}
\right)  -e\left(
\begin{array}
[c]{c}%
v_{y}B_{z}\\
-v_{x}B_{z}%
\end{array}
\right)  -m\left(
\begin{array}
[c]{cc}%
\frac{1}{\tau^{\parallel}} & \frac{1}{\tau^{\perp}}\\
-\frac{1}{\tau^{\perp}} & \frac{1}{\tau^{\parallel}}%
\end{array}
\right)  \left(
\begin{array}
[c]{c}%
v_{x}\\
v_{y}%
\end{array}
\right),
%\end{aligned}
\end{equation}
with the conductivity%
\begin{equation}
\sigma_{xx}=\frac{\frac{ne^{2}\tau^{\parallel}}{m}}{1+(\frac{eB}{m}+\frac
{1}{\tau^{\perp}})^{2}\tau^{\parallel2}},\text{ \ }\sigma_{xy}=-\frac
{ne^{2}(eB+\frac{m}{\tau^{\perp}})\frac{\tau^{\parallel2}}{m^{2}}}%
{1+(\frac{eB}{m}+\frac{1}{\tau^{\perp}})^{2}\tau^{\parallel2}},%
\end{equation}
which gives the same result as that in the Boltzmann theory Eq. \ref{con-skew} when considering only the skew scattering part. In Drude model, $m \mathbf{v}/ \tau$ is a resistive force. The physical meaning of the anisotropic resistive force is that the direction of the force is no longer the same with that of the velocity. This anisotropic force in the Drude theory, on the other hand, is equivalent to the anisotropic scattering in the Boltzmann theory. The difference between Boltzmann theory and the Drude phenomenological theory is that the Drude theory cannot give a specific expression of the longitudinal and transverse relaxation time. 

Converting the conductivity into resistivity, we get 
\begin{equation}
\rho_{xx}  =\frac{m}{e^{2} \tau^{\parallel} n}, 
\end{equation}
\begin{equation}
\rho_{xy}  =-\left( \frac{B}{en}+\frac{m}{e^{2} \tau^{\perp} n} \right).
\end{equation}
Based on our theory, from Eq. \ref{tau-para}, we see that the magnetic field dependence of $\tau^{\parallel}$ contribute to the negative magnetoresistance, while $\frac{1}{\tau^{\perp}}$ contribute to the anomalous Hall effect.

\section{Conclusion}

In summary, we have formulated a classical theory for the magnetotransport
in the 2D Lorentz model. This theory takes into account the effects of the
magnetic field on the electron-impurity scattering
using the recipe of the abstraction of the real scattering process in the classical
Boltzmann framework. We find a correction to the Hall
resistivity in the conventional Boltzmann-Drude theory and a negative magnetoresistivity as a parabolic function of magnetic field. The origin of these results has been analyzed. We have also discussed the relevance between our theory and recent simulation and experimental works. Our theory dominates in a dilute impurity system where the correlation effect is negligible. 

\vskip0.8cm
%\begin{acknowledgments}
We acknowledge useful discussions with Liuyang Sun, Liang Dong, Nikolai A. Sinitsyn, Qi Chen, Liang Du, Fengcheng Wu and Huaiming Guo. Q.N. is supported by DOE (DE-FG03-02ER45958, 
Division of Materials Science and Engineering) in the formulation of our theoy. J.F., C.X. and Y.G. are supported by NSF (EFMA- 1641101) and Welch Foundation (F-1255). 

%\end{acknowledgments}

\vskip0.8cm
\appendix
\label{appendix}

\section{\\Derivation of differential cross section}
\label{APP-A}

\begin{figure}[h]
\setlength{\abovecaptionskip}{0pt}
\setlength{\belowcaptionskip}{0pt}
\scalebox{0.36}{\includegraphics*{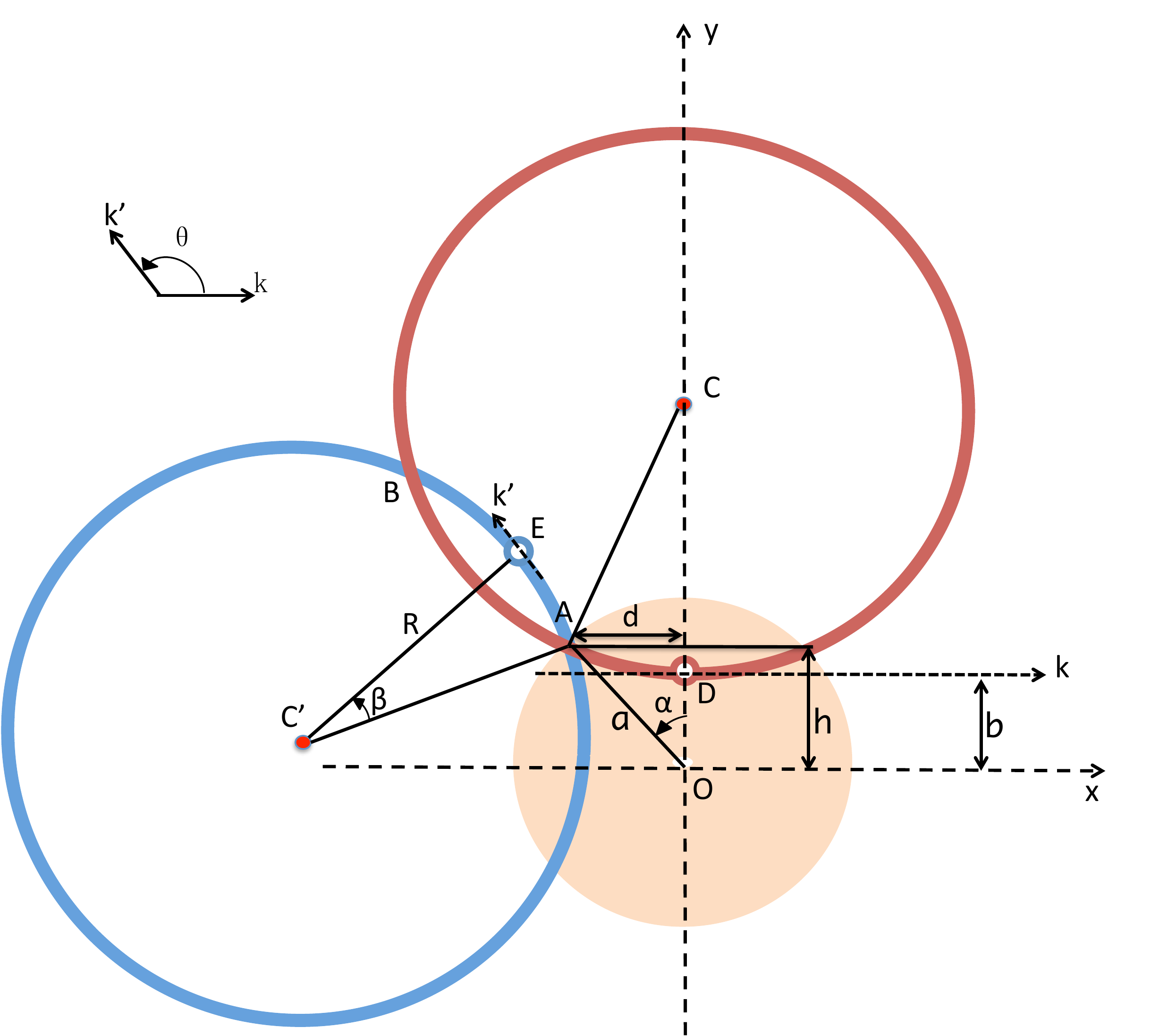}}
\caption{The geometry of the scattering process in the presence of magnetic field. Point $O$ is the center of impurity. Point $C$ and $C'$ are the center of the circle of the initial asymptote, and the circle of the final asymptote, respectively. Point $A$ is the impact point on the scatterer boundary. Point $D$ is the closest approach to the scatterer on the initial asymptote. Point $E$ on the final asymptote is simultaneous with point $D$.}
\label{fig_geometry}
\end{figure}

The relation between impact parameter $b$ and the scattering angle $\theta$ in Eq. \ref{b} is derived from the geometry of the scattering process shown in Fig.~\ref{fig_geometry}. The scattering angle $\theta$ is from $\mathbf{k}$ to $\mathbf{k'}$. Because the direction of $\mathbf{y}$-axis is $\frac{\pi}{2}$ larger than $\mathbf{k}$, and the direction of vector $\overrightarrow{EC'}$ is $\frac{\pi}{2}$ larger than $\mathbf{k'}$, the angle between vector $\overrightarrow{EC'}$ and the $\mathbf{y}$-axis is equal to the scattering angle $\theta$. Equivalently, the angle between $\overrightarrow{AC}$ and $\overrightarrow{AC'}$ is equivalent to the scattering angle $\theta$. Therefore, there is a relation
\begin{equation}
\frac{\theta}{2}=\pi - \angle{OAC}. 
\end{equation}
Along with the cosine theorem for the $\angle OAC$: $a^{2}+R^{2}-2aR\cos \angle OAC=(b+R)^{2}$, there is 
\begin{equation}
a^{2}+R^{2}+2aR\cos\frac{\theta}{2}=(b+R)^{2}. 
\end{equation}
 
Therefore, the expression of $b$ in terms of $\theta$ is
\begin{equation}
b=-R+\sqrt{a^{2}+R^{2}+2aR\cos\frac{\theta}{2}}. 
\label{bk'k_appen}
\end{equation}

For the inverse process $\mathbf{k}^{\prime}\rightarrow \mathbf{k}$, the scattering angle is $2 \pi-\theta$, which falls in the range of $[0, 2 \pi]$. We only focus on the calculation within $[0, 2\pi]$, because the scattering has periodicity of $2\pi$. Beyond $[0, 2\pi]$, the scattering process is the same in every $2\pi$ period. Substituting $\theta$ for $2 \pi-\theta$ in Eq. \ref{bk'k_appen}, we get the differential cross section for $\mathbf{k}^{\prime}\rightarrow \mathbf{k}$ process
\begin{equation}
b=-R+\sqrt{a^{2}+R^{2}-2aR\cos\frac{\theta}{2}}.
\label{bkk'_appen}
\end{equation}

As we see, the two process $\mathbf{k}^{\prime}\rightarrow \mathbf{k}$ and $\mathbf{k} \rightarrow \mathbf{k}^{\prime}$ have different differential cross section, which leads to skew scattering.

\section{\\Derivation of coordinate jump}
\label{APP-B}

The coordinate jump can be derived by $\delta\mathbf{r}=\overrightarrow{OE}-\overrightarrow{OD}$, where $D$ and $E$ are the starting point and ending point (Fig.~\ref{fig_geometry}). $\overrightarrow{OE}=\overrightarrow{OC'}+\overrightarrow{C'E}$. $\overrightarrow{OC'}=-(R+b)\sin(2\alpha)\mathbf{\hat{x}}+ (R+b)\cos(2\alpha)\mathbf{\hat{y}}$. $\overrightarrow{C'E}=R\sin(\theta)\mathbf{\hat{x}} -R\cos(\theta)\mathbf{\hat{y}}$. So $\overrightarrow{OE}=\left[-(R+b)\sin(2\alpha)+R\sin(\theta)\right]\mathbf{\hat{x}}+\left[(R+b)\cos(2\alpha)-R\cos(\theta)\right]\mathbf{\hat{y}}$. 

Also, because $\overrightarrow{OD}=b\mathbf{\hat{y}}$, there are
\begin{align}
\delta\mathbf{x}=\left[-(R+b)\sin(2\alpha)+R\sin(\theta)\right]\mathbf{\hat{x}}, \\
\delta\mathbf{y}=\left[(R+b)\cos(2\alpha)-R\cos(\theta)-b\right]\mathbf{\hat{y}}. 
\end{align}
To replace $b$ and $\alpha$ in terms of $R, a, \theta$, we get
\begin{align}
\delta\mathbf{x}=R\left[  \sin\theta
-\frac{\sin\theta+2\frac{a}{R}\sin\left(  \frac{\theta}{2}\right)  }%
{\sqrt{1+\frac{2a}{R}\cos(\frac{\theta}{2})+\frac{a^{2}}{R^{2}}}}\right]\mathbf{\hat{x}}, \\
\delta\mathbf{y}=2R\sin^{2}\left(
\frac{\theta}{2}\right)  \left[  1-\frac{1}{\sqrt{1+\frac{2a}{R}\cos
(\frac{\theta}{2})+\frac{a^{2}}{R^{2}}}}\right]\mathbf{\hat{y}}.
\end{align}

\section{\\Rigorous treatment of Boltzmann equation}
\label{APP-C}

In general, the Boltzmann equation can be written as%
\begin{widetext}
\begin{equation}
\left(  -e\right)  \left(  \mathbf{E+v}\times\mathbf{B}\right)  \cdot
\frac{\partial f_{\mathbf{k}}}{\hbar\partial\mathbf{k}}=- \int_{0}^{2\pi
}d\theta \left [ W_{\mathbf{kk}^{\prime}}  f\left(  \epsilon
,\mathbf{k}\right) (1-f\left(  \epsilon
,\mathbf{k}^{\prime}\right)) -W_{\mathbf{k}^{\prime}\mathbf{k}} f\left(  \epsilon,\mathbf{k}^{\prime}\right) (1-f\left(  \epsilon
,\mathbf{k}\right))
\right ]  ,
\end{equation}
\end{widetext}
where $W_{\mathbf{kk}^{\prime}}$ is the probability of scattering from $k$ to $k^{\prime}$. 

In a more rigorous treatment shown by Kohn and Luttinger in Eq. (21) of Ref. \cite{KohnLuttinger}, it makes a correspondence between the classical distribution function and the quantum mechanical density matrix. In this treatment, the right hand side of Boltzmann equation becomes 
\begin{widetext}
\begin{equation}
\left(  -e\right)  \left(  \mathbf{E+v}\times\mathbf{B}\right)  \cdot
\frac{\partial f_{\mathbf{k}}}{\hbar\partial\mathbf{k}}=- \int_{0}^{2\pi
}d\theta \left [ W_{\mathbf{kk}^{\prime}}  f\left(  \epsilon
,\mathbf{k}\right) -W_{\mathbf{k}^{\prime}\mathbf{k}} f\left(  \epsilon,\mathbf{k}^{\prime}\right) 
\right ] .
\end{equation}
\end{widetext}
This rigorous treatment in Boltzmann equation has been applied by previous literatures (\cite{Hua2015}, \cite{Sandaram1999}, etc.).

\section{\\The contribution of differential cross section to relaxation time}
\label{APP-D}

Generally, we will prove that $\tau^{\parallel}$ is contributed purely by $\Omega^{S}$, and $\tau^{\perp}$ is contributed purely by $\Omega^{A}$. We first express the differential cross section as a function of scattering angle $f(\theta)$. Then, $\Omega^{A}=\frac{1}{2}(f(\theta)-f(2\pi-\theta))$, $\Omega^{S}=\frac{1}{2}(f(\theta)+f(2\pi-\theta))$. We prove that the term $\Omega^{A} (1+\cos (  \theta ))$ in Eq. \ref{tau-para} and $\Omega^{S} \sin\left(  \theta \right)$ in Eq. \ref{tau-perp} vanishes after integral by
\begin{eqnarray}
&& \int_{0}^{2\pi}d\theta \Omega^{A} (1+\cos (  \theta )) \nonumber \\
&=&\frac{1}{2} \int_{0}^{2\pi}d\theta (f(\theta)-f(2 \pi- \theta) ) (1+\cos (  \theta ))\nonumber \\
&=&\frac{1}{2} \int_{0}^{2\pi}d\theta f(\theta) (1+\cos (  \theta )) \\
&&-\frac{1}{2} \int_{2\pi}^{0}d(2\pi- \theta) f(\theta) \sin (1+\cos ( 2\pi- \theta ))\nonumber \\
&=& 0, \nonumber 
\end{eqnarray}
and
\begin{eqnarray}
&& \int_{0}^{2\pi}d\theta \Omega^{S} \sin\left(  \theta \right) \nonumber \\
&=&\frac{1}{2} \int_{0}^{2\pi}d\theta (f(\theta)+f(2\pi-\theta)) \sin\left(  \theta \right) \nonumber \\
&=&\frac{1}{2} \int_{0}^{2\pi}d\theta f(\theta) \sin\left(  \theta \right)\\
&&+\frac{1}{2} \int_{2\pi}^{0}d(2\pi-\theta) f(\theta) \sin\left(  2\pi-\theta \right)\nonumber \\
&=&0.  \nonumber 
\end{eqnarray}
Therefore, we have demonstrated that the symmetric and antisymmetric part of the probability of scattering contribute to the conventional transport relaxation time $\tau^{\parallel}$ and the transverse relaxation time $\tau^{\perp}$, respectively.

\section{\\Application to soft potential under magnetic field}
\label{APP-E}

\begin{figure}[h]
\setlength{\abovecaptionskip}{0pt}
\setlength{\belowcaptionskip}{0pt}
\scalebox{0.32}{\includegraphics*{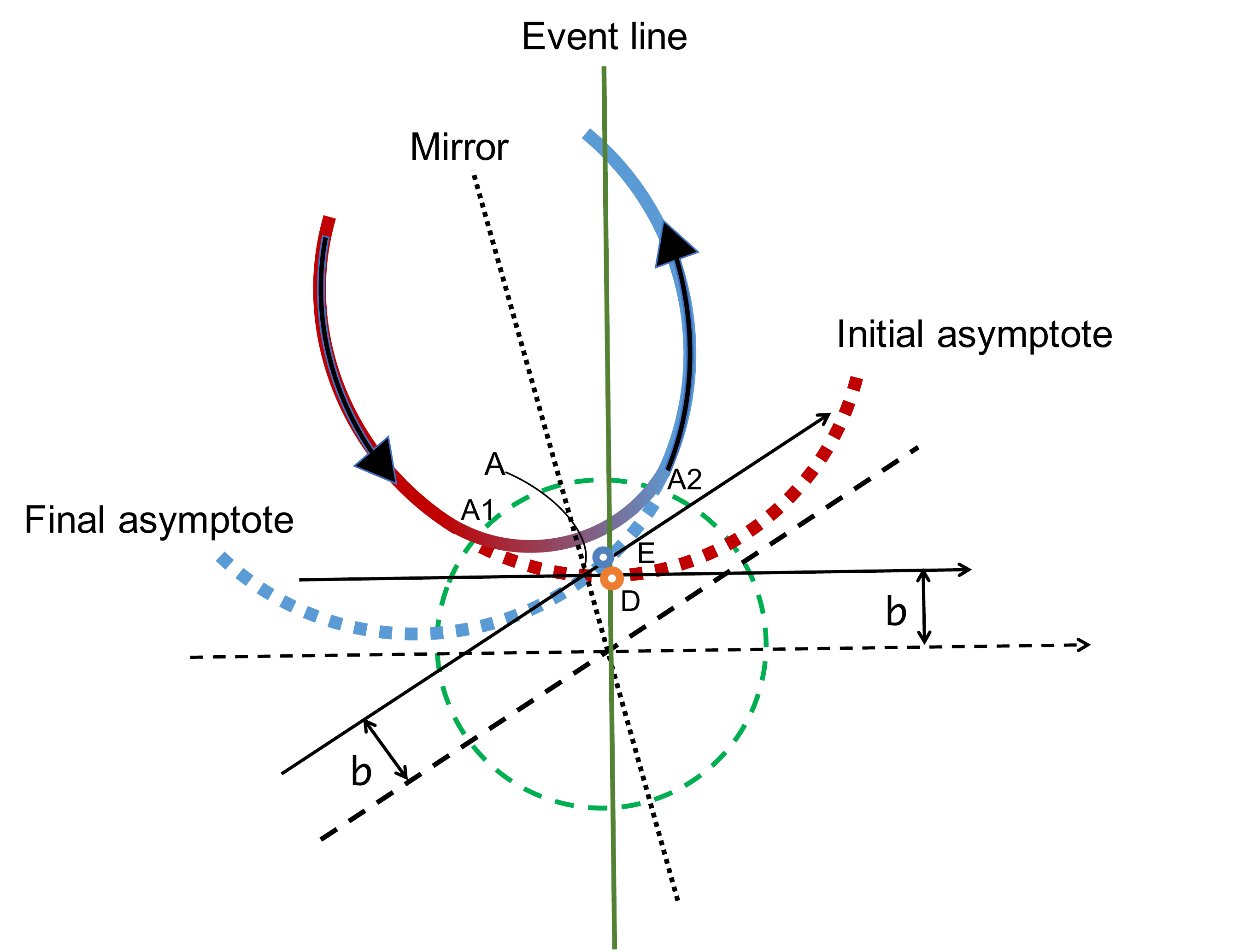}}
\caption{The plot of soft potential scattering. The green dashed circle is the range, within which the potential influences on. The real trajectory, incoming with red solid curve and outgoing with blue solid curve, intersects the green dashed circle at two points $A1$ and $A2$. The red dashed curve and blue dashed curve are the initial asymptote and final asymptote, respectively.  $A$ is the crossing point between the red and blue dashed curve. The incoming trajectory and the outgoing trajectory are symmetric about the mirror line, which passes through point $A$ and the scatterer center. The red dot $D$ is the starting point on the event line, and the blue dot $E$ is the ending point that is at the same time with the event line.}
\label{fig_soft_scatterer}
\end{figure}

We here demonstrate how to apply our abstraction method to the soft potential scattering in magnetic field. We assume a central soft potential with a well-defined center of scattering. The soft potential goes to zero at infinity. As a result, we can draw a circle (green dashed circle) around the potential center that is large enough so that the interaction between electron and potential outside the circle is negligible (See Fig. \ref{fig_soft_scatterer}). The real trajectory intersects the green dashed circle at two points $A1$ and $A2$. Before reaching $A1$ and after reaching $A2$, the real trajectory coincides with the initial and final asymptote, respectively. We then extend the initial and final asymptote after $A1$ and before $A2$ to their intersecting point $A$. The time from $A$ to $A1$ and $A$ to $A2$ are the same, due to the mirror symmetry. The mirror symmetry is proven by the following. The angular momentum outside the green circle $\mathbf{L}$ satisfies
\begin{eqnarray}
\frac{d\mathbf{L}}{dt}&=&\mathbf{r} \times m \dot{\mathbf{v}} \nonumber \\
&=& \mathbf{r} \times (-e \dot{\mathbf{r}} \times \mathbf{B}) \nonumber \\
&=& -e (\mathbf{r}  \cdot  \dot{\mathbf{r}} ) \mathbf{B}  \\
&=& -\frac{d}{dt} (\frac{1}{2} e r^2 \mathbf{B}).  \nonumber \\
\end{eqnarray}
Therefore, the quantity $\mathbf{r} \times m \mathbf{v}+\frac{1}{2} e r^2 \mathbf{B} = constant$. The $\frac{1}{2} e r^2 \mathbf{B}$ is constant on the green circle, because $|r_1|=|r_2|=R$. Therefore, the angular momentum $\mathbf{r} \times m \mathbf{v}$ is constant at $A_1$ and $A_2$. Due to energy conservation, $|\mathbf{v_1}|=|\mathbf{v_2}|$. Therefore, the velocity at $A_1$ and $A_2$ are symmetric (if reversing the velocity at $A_2$ by $\pi$) about the mirror line which passes through the intersection of the elongated velocity of $A_1$ and $A_2$. The intersection of the initial and final asymptote $A$ falls on the mirror line as well. 

We can define the starting point and ending point in a similar fashion. The starting point, marked as $D$ (red dot) in Fig. \ref{fig_soft_scatterer}, is the closest approach to the center of scatterer on the initial asymptote. The event line, connecting the starting point and the center of the scatterer, marks the occurring of scattering event. The blue dot on the final asymptote is the ending point $E$, which satisfies that the time from $A$ to $D$ and $A$ to $E$ are the same. By definition, both $D$ and $E$ are on the asymptote, not on real trajectory. Therefore, the ending point $E$ is not influenced by the scattering potential. The purpose of our method is to provide a way to appropriately parametrize the scattering process. 

In summary, the electron motion in a soft potential and magnetic field can be abstracted as follows. The electron moves along the initial asymptote to the starting point, gets scattered to the ending point and finally moves away from the scatterer along the final asymptote. The initial asymptote, final asymptote and the event line have the same definition with those defined in the manuscript with the hard disk potential.

\begin{figure}[h]
\setlength{\abovecaptionskip}{0pt}
\setlength{\belowcaptionskip}{0pt}
\scalebox{0.17}{\includegraphics*{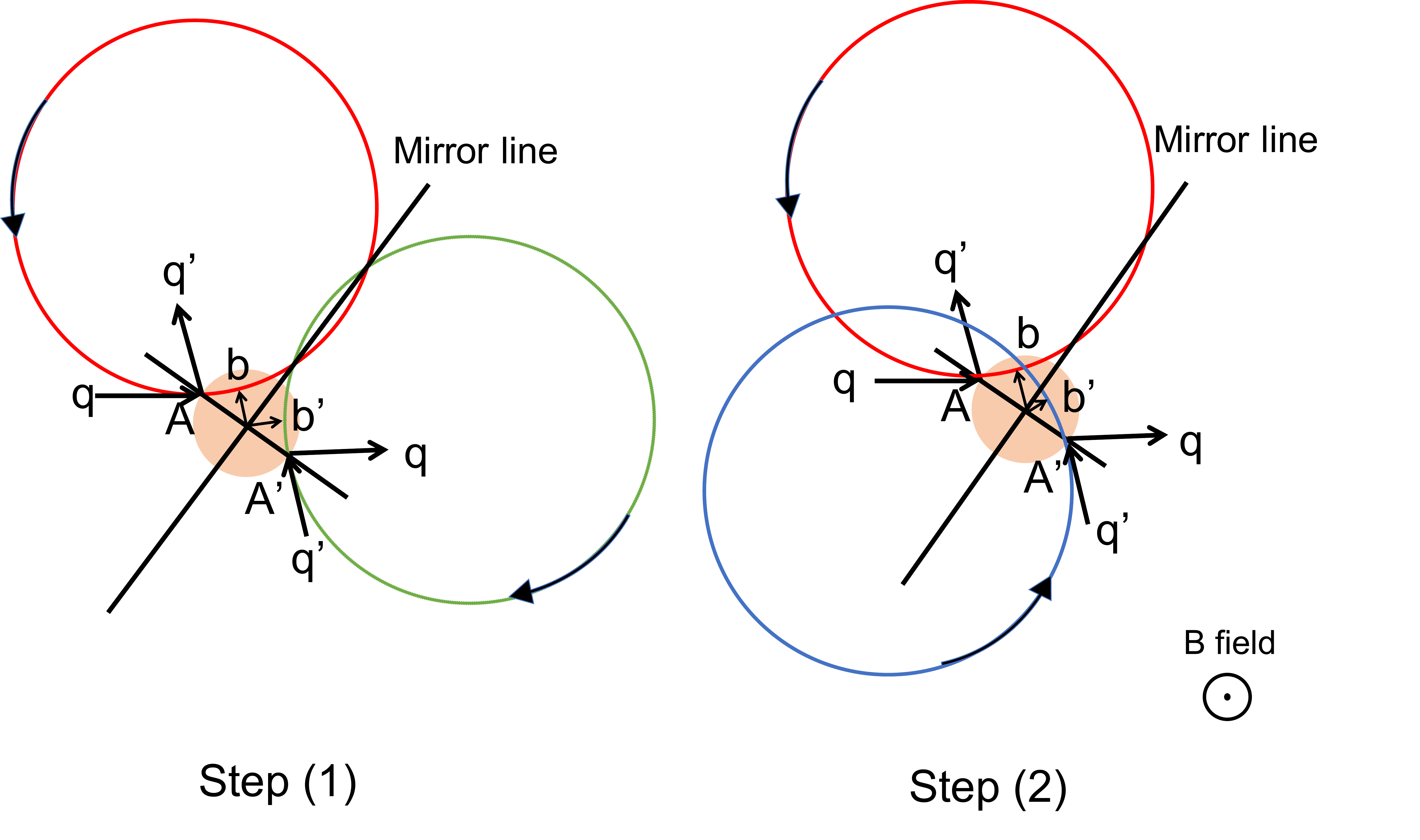}}
\caption{The sketch of scattering process from $\mathbf{q} \rightarrow \mathbf{q}^{\prime}$ and $\mathbf{q}^{\prime} \rightarrow \mathbf{q}$ process: (1) with symmetric scattering angle $\theta$  and symmetric impact parameter $b=b'$ due to reversed magnetic field; (2) with symmetric scattering angle $\theta$ but different impact parameter $b \neq b'$ due to correct magnetic field direction. The left side of mirror line plots the $\mathbf{q} \rightarrow \mathbf{q}^{\prime}$ process, and the right side of mirror line plots the $\mathbf{q}^{\prime} \rightarrow \mathbf{q}$ process. $A$ and $A'$ are the impact point at each side. $b$ and $b'$ are the impact parameter at each side. In step (1) with reversed magnetic field direction, the $\mathbf{q} \rightarrow \mathbf{q}^{\prime}$ and $\mathbf{q}^{\prime} \rightarrow \mathbf{q}$ process are symmetric with respect to the mirror line. Step (2) plots $\mathbf{q}^{\prime} \rightarrow \mathbf{q}$ process with correct magnetic field direction. The impact parameter $b'$ is no longer the same with $b$. }
\label{fig_geo_OO'}
\end{figure}

\section{\\Alternative understanding of anisotropic scattering}
\label{APP-F} 

We can use the impact point on the boundary of scatterer to understand how differential cross section is different between $\mathbf{q} \rightarrow \mathbf{q}^{\prime}$ and $\mathbf{q}^{\prime} \rightarrow \mathbf{q}$ process, using the following two steps as shown in Fig. \ref{fig_geo_OO'} (where $\mathbf{q}$ and $\mathbf{q}^{\prime}$ are the momentum on the scatterer boundary). (1): We first draw the incoming trajectory for $\mathbf{q} \rightarrow \mathbf{q}^{\prime}$ process. We then draw a line crossing the center of scatterer and parallel to the tangent line at the impact point A as mirror line. We then draw the mirror image of the $\mathbf{q} \rightarrow \mathbf{q}^{\prime}$ process at right side of the mirror line (Fig. \ref{fig_geo_OO'}, step (1)). The impact point goes to A' which is the mirror image of A. Now the impact parameter is the same for the two processes, i.e. $b=b'$, which gives no skew scattering. (2): The magnetic field direction determines that the rotation of electron should be counterclockwise. However, upon doing step (1), the rotation is clockwise in $\mathbf{q}^{\prime} \rightarrow \mathbf{q}$ process, which is incorrect. The correct way for $\mathbf{q}^{\prime} \rightarrow \mathbf{q}$ is to draw the counterclockwise trajectory tangential to the incident and scattered momentum at A' (we only show the incident trajectory in Fig. \ref{fig_geo_OO'}, step (2)). It is clear that the impact parameter $b'$ is different from $b$, which generates asymmetry between  $\mathbf{q} \rightarrow \mathbf{q}^{\prime}$ and $\mathbf{q}^{\prime} \rightarrow \mathbf{q}$ process. While the scattering angle stays the same, the differential cross section is different due to $\Omega=|db/d\theta|$. This analysis also shows the role of the magnetic field in the skew scattering. The $\mathbf{q}^{\prime} \rightarrow \mathbf{q}$ process in step (1) reverses the direction of the magnetic field. In order to correct this, the electron trajectory has to be subject to another mirror/time reversal operation as in step (2). The resulting correct $\mathbf{q}^{\prime} \rightarrow \mathbf{q}$ process is thus no longer symmetric with the original $\mathbf{q} \rightarrow \mathbf{q}^{\prime}$ process.

\bibliographystyle{apsrev4-1}

%\bibliography{reference}

\end{document}